\documentclass{article}
\usepackage{jcappub}
\usepackage{graphicx} 
\usepackage{bm}
\usepackage{amsmath}
\usepackage{amsmath,amsfonts,amssymb}
\usepackage{mathtools}
\usepackage{braket}
\usepackage{setspace}
\usepackage{hyperref}
\hypersetup{
    colorlinks=true, 
    linkcolor=blue,  
    urlcolor=blue 
  }
\usepackage{doi}

\usepackage{xcolor}

\usepackage{soul}
\usepackage[normalem]{ulem}
\usepackage{cancel}

\title{Cosmic Birefringence from \\ Large Lepton Asymmetry}
\author[a,b]{Kentaro Kasai}

\author[b,c,d]{and Ippei Obata}
\affiliation[a]{ICRR, University of Tokyo, Kashiwa 277-8582, Japan}
\affiliation[b]{Theory Center, Institute of Particle and Nuclear Studies (IPNS), High Energy Accelerator Research Organization (KEK), 1-1 Oho, Tsukuba, Ibaraki 305-0801, Japan}
\affiliation[c]{Department of Physics, Rikkyo University, Toshima, Tokyo 171-8501, Japan}
\affiliation[d]{Kavli Institute for the Physics and Mathematics of the Universe (Kavli IPMU), WPI, UTIAS, The University of Tokyo, Kashiwa, Chiba 277-8568, Japan}

\emailAdd{kkasai98@post.kek.jp}
\emailAdd{iobata@rikkyo.ac.jp}

\abstract{Cosmic birefringence, the rotation of the polarization plane of cosmic microwave background (CMB) photons, provides an intriguing probe of physics beyond the Standard Model. In this paper, we investigate a vector Chern-Simons coupling between the neutrino current and photons as a possible origin of isotropic cosmic birefringence. Firstly, we perform the first analysis of anisotropic cosmic birefringence induced by the neutrino current. We find that the anisotropic signal sourced by lepton-number isocurvature fluctuations can be important on large scales, whereas on small scales, the anisotropic signal induced by gravitational effects becomes dominant comparing with the axion-like particle (ALP) scenario. Secondly, we show that a large lepton asymmetry generated through the Affleck-Dine mechanism can successfully account for the observed isotropic birefringence, and discuss its possible connection to the recently reported low helium-4 abundances in Subaru EMPRESS and Atacama Cosmology Telescope (ACT).}

\date{September 2025}

\begin{document}

\maketitle

\section{Introduction}

Testing the parity violation is one of the essential ideas to understand fundamental physics.
In cosmology, cosmic birefringence \cite{Carroll:1989vb,Harari:1992ea,Carroll:1991zs,Nodland:1997cc,Komatsu:2022nvu}, a cosmological polarization rotation effect in the presence of parity-violating physics, has been received a lot of attention.
Parity-violating cosmological background produces the parity-odd correlation function in the cosmic microwave background (CMB) map \cite{Lue:1998mq,Feng:2006dp,Liu:2006uh}.
While the finite polarization rotation angle has been measured in CMB data \cite{Feng:2006dp,WMAP:2010qai,Planck:2016soo,Namikawa:2020ffr,SPT:2020cxx}, these results did not consider or suffered from the effect of systematic uncertainty.
Recently, however, a new calibration method for the isotropic cosmic birefringence has been developed.
By separating the CMB intrinsic angle from galaxy foreground, due to the huge difference between cosmological distance and galactic distance, assuming cosmic birefringence effect is relevant from the CMB photon emitted at the last scattering surface, we can distinguish the intrinsic rotation angle from miscalibration angle \cite{Minami:2019ruj}.
By using these strategies, a nonzero rotation angle over the 2 sigma has been reported from the Planck PR3 data \cite{Minami:2020odp}.
The subsequent Planck PR4 studies improved the statistical significance \cite{Diego-Palazuelos:2022dsq,Remazeilles:2025wzd}, and the latest joint analysis of Planck/WMAP data tells us the presence of isotropic cosmic birefringence angle for 3.6 $\sigma$ \cite{Eskilt:2022cff}.
The subsequent results from Atacama Cosmology Telescope (ACT) and South Pole Telescope (SPT) may also support a presence of birefringence signal, while they used the different calibration method and systematic effects should be still considerable \cite{AtacamaCosmologyTelescope:2025blo,Diego-Palazuelos:2025dmh}.
While there are some debates on the impact of systematics such as optical/instrumental physics or astrophysical foreground \cite{Diego-Palazuelos:2022dsq,Clark:2021kze,Abghari:2022bet,Cukierman:2022kei,Diego-Palazuelos:2022cnh,Monelli:2022pru,Cosmoglobe:2023pgf,Taurus:2024xre,Hervias-Caimapo:2024ili,Sullivan:2025bva, Lonappan:2026unb}, future CMB telescopes such as Simons Observatory or LiteBIRD are expected to improve the sensitivity and potentially confirm the signal \cite{SimonsObservatory:2018koc,Abitbol:2020fvn,Jost:2022oab,SimonsObservatory:2025wwn,LiteBIRD:2025yfb,Dou:2024wdy,Dou:2025luz}.

The measured angle is consistent with frequency-independent \cite{Eskilt:2022wav}, which is hard to explain within the Standard Model \cite{Nakai:2023zdr}.
One of the promising scenario is to consider the photon coupling to the axion like particles (ALPs) \cite{Svrcek_2006,Arvanitaki:2009fg}.
ALP background differentiates the phase velocity of circular polarized photon, making the polarization orientation whose rotation angle magnitude depends on the ALP photon coupling constant and ALP field displacement between CMB epoch and present time.
To explain the isotropic cosmic birefringence signal, a variety of cosmological ALP model \cite{Fujita:2020aqt,Fujita:2020ecn,Fujita:2022qlk,Choi:2021aze,Lin:2022niw,Lin:2022khg,Liang:2024vnd,Girmohanta:2023ghm,Gasparotto:2022uqo,Fung:2021wbz,Fung:2021fcj,Luu:2021yhl,Takahashi:2020tqv,Kitajima:2022jzz,Gonzalez:2022mcx,Ferreira:2023jbu,Lee:2024xjb,Lee:2025yvn,Jain:2021shf,Jain:2022jrp,Nakagawa:2021nme,Obata:2021nql,Murai:2024yul,Zhang:2024dmi,Shao:2025jhp,Mehta:2021pwf,Reig:2021ipa,Alvey:2021hjp,Gasparotto:2023psh,Choi:2023gin,Gendler:2023kjt,Sheridan:2024vtt,Agrawal:2024ejr} and their dedicated observational implications \cite{Sherwin:2021vgb,Nakatsuka:2022epj,Galaverni:2023zhv,Namikawa:2025sft,Ballardini:2025apf,Carralot:2026kps,Murai:2022zur,Eskilt:2023nxm,Yin:2023srb,Kochappan:2024jyf,Yin:2025fmj,Yin:2026gss,Hotinli:2022wbk,Lee:2022udm,Namikawa:2023zux,Naokawa:2023upt,IdicherianLonappan:2025trj,Yin:2024fez,Naokawa:2025shr,Naokawa:2024xhn,Namikawa:2025doa,Tada:2024znt,Berbig:2024aee,Nakagawa:2025ejs,Lin:2025gne,Barman:2025ryg,Yamashita:2021gro} have been developed (also, see recent reviews \cite{Komatsu:2022nvu,Hui:2021tkt,Perivolaropoulos:2021jda,Abdalla:2022yfr,OHare:2024nmr,Arza:2026rsl}).
Or, nontrivial topology of our universe may produce the parity-odd CMB corelation \cite{COMPACT:2024cud}. Furthermore, it is indicated that the evolution of 3-form dark energy which couples to photons can explain the origin of isotropic cosmic birefringence~\cite{Kamada:2026vbw}.

In this paper, we focus on an alternative scenario to the ALPs as a source of the cosmic birefringence. 
Indeed, original works~\cite{Carroll:1989vb,Carroll:1991zs} show that an interaction of a vector current couples to the vector-like Chern-Simons operator. 
One can rewrite it into the scalar Chern-Simons operator where a vector current corresponds with a gradient of ALP field.
Otherwise, unless the vector current is a constant and violates the Lorentz symmetry in spacetime \cite{Caloni:2022kwp,Nilsson:2023sxz}, it explicitly depends on a gauge potential and violates the gauge symmetry.
Therefore, alternative auxiliary field, such as Stueckelberg trick, is necessary to compensate for a gauge symmetry.
As a medium, to explain the isotropic cosmic birefringence, it should be homogeneous in our universe, and would be a neutral particle since charged particle's abundance is much suppressed.
One of such a candidate is a cosmic neutrino background \cite{Geng:2007va,Ho:2010aq,Zhou:2023aqz}.
To explain the cosmic birefringence, it requires a large lepton asymmetry comparing to the baryon asymmetry in the early universe. This possibility has another interesting observational motivation, that is, the low primordial Helium-4 abundance suggested by EMPRESS~\cite{Matsumoto:2022tlr, Yanagisawa:2025mgx} and ACT~\cite{AtacamaCosmologyTelescope:2025nti}.
It's potentially interesting to study those two implications for the parity and charge asymmetry in the universe from the fundamental physics. 

In this paper, we explore a connection between a concrete UV-inspired model for the origin of large lepton asymmetry and the signal of cosmic birefringence.
In general, if a large lepton asymmetry is generated in the epoch before the electroweak phase transition, baryon number with a same order is produced via the electroweak sphaleron process~\cite{Kuzmin:1985mm}, which spoils the success of Big Bang nucleosynthesis (BBN). To avoid this problem, one should consider an asymmetry generation well after the electroweak phase transition or some mechanism which suppresses the sphaleron process. Here, we focus on the Affleck-Dine mechanism~\cite{Affleck:1984fy,Dine:1995kz}, which can realize a large lepton asymmetry without overproduction of baryon number~\cite{Kawasaki:2002hq,Kawasaki:2022hvx,Kasai:2024diy,Akita:2025zvq}~\footnote{Recently, it is indicated that a large lepton asymmetry can also be generated through the helicity of the $U(1)_{L_\mu-L_\tau}$ gauge field, which is produced by the dynamics of an axion field during inflation.~\cite{Hu:2025nep}.}. In this model, a CP-violating dynamics of a leptonic complex scalar field, which naturally appears in minimal supersymmetric Standard Model (MSSM), generates a lepton asymmetry. Notably, the problem of baryon number overproduction is successfully overcome if we consider a formation of classically stable non-topological soliton of the scalar field, known as Q-balls~\cite{Kusenko:1997zq,Enqvist:1997si, Enqvist:2003gh,Kawasaki:2002hq, Kawasaki:2022hvx}. Sphaleron process is largely suppressed inside the Q-ball, since a large VEV of the complex scalar breaks the electroweak symmetry even before the epoch of electroweak phase transition. However, the Q-balls can decay into SM particles such as neutrinos via quantum effect, which typically takes place well after the electroweak phase transition, enabling us to successfully explain the origin of a lepton asymmetric SM plasma before BBN. We show that, within the gauge-mediated supersymmetry (SUSY) breaking scenario, a wide range of the parameter space can realize the generation of large lepton asymmetry, which is compatible for explaining the origin of isotropic cosmic birefringence. We also show that some parameter sets can also explain the lepton asymmetry suggested by recent EMPRESS results~\cite{Matsumoto:2022tlr,Yanagisawa:2025mgx}.

Furthermore, we also focus on the anisotropic cosmic birefringence as a powerful tool to distinguish our model from other scenarios.
For ALP scenarios, anisotropic cosmic birefringence is provided with the ALP fluctuations \cite{Pospelov:2008gg,Li:2008tma,Kamionkowski:2008fp,Caldwell:2011pu,Greco:2022ufo,Greco:2022xwj,Greco:2024oie,Cai:2022zad} and it has not yet been measured but has an upper bound on the magnitude of two point correlation function of the rotation angle \cite{Gluscevic:2012me,POLARBEAR:2015ktq,BICEP2:2017lpa,Contreras:2017sgi,Namikawa:2020ffr,SPT:2020cxx,Gruppuso:2020kfy,Pogosian:2019jbt}.
One of differences is scale dependence of power spectrum depending on the ALP scenario, for example, ALP domain wall network provides a drastic scale-dependent power spectrum \cite{Kitajima:2022jzz,Gonzalez:2022mcx,Ferreira:2023jbu} comparing with the ALP alignment mechanism scenario, which predicts a nearly scale invariant spectrum sourced by a curvature perturbation on large scales.
Therefore, developing a detailed study on the anisotropic cosmic birefringence is important to identify the cosmological scenario~\footnote{Recently, the redshift evolution of isotropic cosmic birefringence has been investigated as a means of distinguishing neutrino-induced case and the ALP case through the CMB $EB$ power spectrum~\cite{Yin:2026qlh}.}.

This paper is organized as follows. 
In section~\ref{sec: Model setup}, we present a model of generating cosmic birefringence from cosmic neutrino background via the vector Chern-Simons interactions to photon. 
In section~\ref{sec: ansotropic CB}, we evaluate the anisotropic cosmic birefringence sourced by lepton isocurvature perturbation. 
In section~\ref{sec: AD mechanism as an origin}, as an example we consider Q-ball scenario and find a parameter search for the successful scenario. 
In section~\ref{sec: summary and discussions}, we discuss and conclude our work.
Throughout this paper, we set a natural unit $\hbar=c=1$.

\section{Model Setup}
\label{sec: Model setup}

In this section, we present a model which predicts a cosmic birefringence by the neutrino via a vector-CS interaction to photon, following Ref.~\cite{Zhou:2023aqz}. 
The interaction Lagrangian density is as follows:
\begin{equation}
\mathcal{L} \supset -\dfrac{1}{2}\dfrac{c_\nu}{M^2}J_\mu A_\nu \tilde{F}^{\mu\nu} \ ,
\label{eq : vector CS coupling}
\end{equation}
where the vector four-current of Dirac neutrino $\nu$ reads
\begin{equation}
J_\mu = \bar{\nu}\gamma_\mu\nu \ .
\end{equation}
For simplicity, we consider the interaction with only the electron neutrino: $\nu = \nu_e$.
Then, in a rest frame only $J_0$ component appears and it is given by the difference of neutrino and anti-neutrino number density:
\begin{equation}
J_0 = \Delta n_\nu \equiv n_\nu - n_{\bar{\nu}} \ .
\end{equation}

In the presence of neutrino-photon coupling, the orientation of polarization plane rotates, which is characterized by the following birefringence angle:
\begin{equation}
\beta(t,\hat{\bm{n}}) = \dfrac{c_\nu}{2M^2}\int_e^0 J_0(t,\hat{\bm{n}})
{\rm d}t = \dfrac{c_\nu}{2M^2}\int_e^0 \Delta n_\nu(t,\hat{\bm{n}}) 
{\rm d}t \ ,
\label{eq : formula for the birefringence angle}
\end{equation}
where $e$ denotes a position where polarization was emitted and $0$ denotes at present.
We separate $\beta$ into its isotropic part and anisotropic (spatial-dependent) part:
\begin{equation}
\beta(t,\hat{\bm{n}}) = \bar{\beta}(t) + \delta\beta(t,\hat{\bm{n}}) \ .
\end{equation}
First, let us evaluate the isotropic component of birefringence angle $\bar{\beta}$.
The background number density is given by an integration of distribution function in momentum space:
\begin{equation}
\bar{n}_\nu(t) = \dfrac{g_s}{a^3}\int\dfrac{{\rm d}\bm{q}}{(2\pi)^3}\bar{f}_{\nu}(q,t) \ ,
\end{equation}
where $g_s=1$ is a number degree of freedom of left-handed electric neutrino and $\bar{f}_{\nu}$ is an unperturbed distribution function at the thermal equilibrium.
The neutrino obeys the Fermi-Dirac statistics and its phase space distribution function at thermal equilibrium is given by
\begin{equation}
\bar{f}_{\nu}(q, t) = \dfrac{1}{e^{(\epsilon(q,t)-a\mu_\nu)/(aT_\nu(a))}+1} \ ,
\end{equation}
where $\epsilon(q,t)=\sqrt{q^2 + a(t)^2m^2}$ with $\bm{q}$ being the comoving momentum and $\mu_\nu$ is a chemical potential of neutrino.
At the neutrino decoupling epoch, neutrino is highly relativistic and we can approximate it as
\begin{equation}
\bar{f}_{\nu}(q,t) \simeq \bar{f}_{\nu}(q) \equiv \dfrac{1}{e^{(q-a\mu_\nu)/(aT_\nu(a))}+1}
\end{equation}
and this function works even after neutrinos become non-relativistic which occurs much after decoupling epoch.
Since $\mu_{\bar{\nu}} = -\mu_\nu$, the difference of neutrino and anti neutrino number density at the relativistic limit is given by
\begin{equation}
\Delta \bar{n}_\nu = 
\dfrac{g_sT_\nu^3}{6\pi^2}(\pi^2\xi_\nu + \xi_\nu^3) \simeq \dfrac{g_sT_\nu^3}{6}\xi_\nu \label{eq: BG}
\end{equation}
in terms of degeneracy parameter $\xi_\nu \equiv \mu_\nu/T_\nu, \ |\xi_\nu| \ll 1$.
In the Friedmann‐Lema\^\i tre‐Robertson-Walker (FLRW) spacetime ${\rm d}s^2 = -{\rm d}t^2 + a(t)^2{\rm d}\bm{x}^2$, assuming the matter-dominated universe and $a(t_0)=1$, we have
\begin{equation}
{\rm d}t = \dfrac{{\rm d}a}{Ha} \ , \qquad H \equiv \dfrac{\dot{a}}{a} \simeq H_0\sqrt{\Omega_m}a^{-3/2} \ ,
\end{equation}
where $H_0 = 100h \ \text{km/(s}\cdot\text{Mpc})$ is a Hubble constant with a dimensionless Hubble parameter $h \sim 0.7$, and $\Omega_m$ is a matter density parameter at present.
Therefore, we obtain
\begin{align}
\bar{\beta} &= \dfrac{c_\nu}{2M^2}\int_e^0 \dfrac{g_sT_\nu^3}{6\pi^2}(\pi^2\xi_\nu + \xi_\nu^3)\dfrac{{\rm d}a}{Ha} \notag \\
&\simeq \dfrac{c_\nu}{2M^2}\int_e^0 \dfrac{g_sT_{\nu,0}^3\xi_\nu}{6a^{5/2}}\dfrac{{\rm d}a}{H_0\sqrt{\Omega_m}} \notag \\
&\simeq \dfrac{c_\nu}{2M^2}\dfrac{g_sT_{\nu,0}^3\xi_\nu}{6H_0\sqrt{\Omega_m}}\dfrac{2}{3}\dfrac{1}{a(t_e)^{3/2}} \qquad (a(t_e) \ll a(t_0))  \ .
\end{align}
Regarding the cutoff scale, it is constrained by the rare $Z$-boson decay at LEP \cite{Zhou:2023aqz}:
\begin{equation}
\dfrac{c_\nu}{M^2} \lesssim \dfrac{1.1\times 10^3}{(\rm TeV)^2} \ .
\label{eq: upper bound of cnu}
\end{equation}
Substituting $T_{\nu,0} = (4/11)^{1/3}T_0 \simeq 1.95$ K, we obtain
\begin{align}
\bar{\beta} \sim 0.3^\circ
{c_\nu}
\left(\frac{\xi_\nu}{0.05}\right)
\left(\frac{10\,\mathrm{TeV}}{M}\right)^2 .
\end{align}

\section{Anisotropic Cosmic Birefringence}
\label{sec: ansotropic CB}

Next, we present the perturbation of neutrino background and derive the expression for the anisotropic cosmic birefringence angular power spectrum.
The fluctuations of cosmic neutrino background are characterized by that with the distribution function:
\begin{equation}
f_\nu(\bm{k}, \bm{q},t) = \bar{f}_\nu(q)(1 + \Psi_\nu(\bm{k}, \bm{q},t)),
\end{equation}
which obey the linearized collision-less Boltzmann equation \cite{Ma:1995ey}:
\begin{equation}
\dfrac{\partial \Psi_\nu}{\partial \tau} + i\dfrac{q}{\epsilon}(\bm{k}\cdot\hat{\bm{q}})\Psi_\nu + \dfrac{{\rm d} \ln \bar{f}_{\nu}(q)}{{\rm d} \ln q}\left[ \dot{\phi} - i\dfrac{\epsilon}{q}(\bm{k}\cdot\hat{\bm{q}})\psi \right] = 0 \ ,
\end{equation}
where $\hat{\bm{q}} \equiv \bm{q}/q$ and $\phi, \psi$ are the metric scalar perturbations in the conformal Newtonian gauge:
\begin{equation}
{\rm d}s^2 = a(\tau)^2\left[ - (1+2\psi){\rm d}\tau^2 + (1 - 2\phi)
{\rm d}\bm{x}^2\right] \ .
\end{equation}
Then, expanding $\Psi_\nu$ by Legendre polynomials
\begin{equation}
\Psi_\nu(\bm{k},\bm{q}, \tau) = \sum_{\ell = 0}^\infty (-i)^\ell(2\ell + 1)\Psi_{\nu,\ell}(\bm{k}, q, \tau) P_\ell(\cos\theta) \ ,
\end{equation}
and in terms of a dimensionless time variable $x \equiv k\tau$, the Boltzmann hierachy equations are obtained:
\begin{align}
\Psi_{\nu,0}' &= -\dfrac{q}{\epsilon}\Psi_{\nu,1} - \phi'\dfrac{{\rm d} \ln \bar{f}}{{\rm d} \ln q} \ , \\
\Psi_{\nu,1}' &= \dfrac{q}{3\epsilon}(\Psi_{\nu,0} - 2\Psi_{\nu,2}) - \dfrac{\epsilon}{3q}\psi\dfrac{{\rm d} \ln \bar{f}}{{\rm d} \ln q} \ , \\
\Psi_{\nu,\ell}' &= \dfrac{q}{(2\ell+1)\epsilon}\left[\ell\Psi_{\nu,\ell-1} - (\ell+1)\Psi_{\nu,\ell+1}\right] \ , \qquad \ell \geq 2 \ .
\end{align}
where the prime means the derivative with $x$.

To analyze the fluctuation of neutrino number density, we derive the solution for the monopole component $\Psi_{\nu,0}$.
Assuming that the initial condition at $x = x_i$ for higher multipole components are negligible: $\Psi_{\nu,\ell\geq3}(x_i \ll 1) \simeq 0$, the explicit solution of the linear order perturbation to the distribution function $\Psi_{\nu,0}$ is given by~\cite{Shoji:2010hm}:
\begin{align}
\Psi_{\nu,0}(k,q,x) = &\Psi_{\nu,0}(k,q,x_i)j_0(z-z_i) - 3\Psi_{\nu,1}(k,q,x_i)j_1(z-z_i) + 5\Psi_{\nu,2}(k,q,x_i)j_2(z-z_i) \notag \\
 &+ \dfrac{{\rm d} \ln \bar{f}_\nu}{{\rm d} \ln q}\left[\int_{x_i}^x dx'\left( \dfrac{\epsilon(q,x')}{q}\psi(k,x')j_1(z-z') - \dfrac{{\rm d}\phi(k,x')}{{\rm d} x}j_0(z-z') \right) \right] \ , \label{eq: solution}
\end{align}
where $x \equiv k\tau$, $j_n$ is the spherical Bessel function, and we have defined the integration variable $z(x)$ as
\begin{equation}
z - z' \equiv \int^x_{x'}\dfrac{q}{\epsilon(q,x'')}
{\rm d}x'' = x - x' \qquad (m_\nu = 0) \ . \label{eq: z}
\end{equation}
In the following, we further ignore the initial condition for higher moments (the second and the third term in right hand side in Eq.~\eqref{eq: solution}), since they are suppressed by the power of $x_i$~\cite{Ma:1995ey,Shoji:2010hm}.

Let us solve the initial condition for $\Psi_\ell$.
At this time, the neutrino is highly relativistic and thereby we can assume $\epsilon \simeq q$. 
By defining $\delta\mu_\nu$ as the initial chemical potential perturbations, we can find it from the expansion of distribution function in terms of $\delta T_\nu$ and $\delta\mu_\nu$:
\begin{align}
f_\nu(q) &= \bar{f}_\nu(q) + \dfrac{e^{(q-a\mu_\nu)/aT_\nu}}{(e^{(q-a\mu_\nu)/aT_\nu}+1)^2}\dfrac{\delta\mu_\nu}{T_\nu} + \dfrac{e^{(q-a\mu_\nu)/aT_\nu}}{(e^{(q-a\mu_\nu)/aT_\nu}+1)^2}\dfrac{q-a\mu_\nu}{aT_\nu}\dfrac{\delta T_\nu}{T_\nu} + \mathcal{O}(\delta \mu_\nu^2, \delta T_\nu^2, \delta\mu_\nu \delta T_\nu) \notag \\
&\simeq \bar{f}_\nu(q)\left[ 1 -\dfrac{\text{d}\ln \bar{f}_\nu(q)}{\text{d}\ln q}\left(\dfrac{a\delta\mu_\nu}{q} + \dfrac{q-a\mu_\nu}{q}\dfrac{\delta T_\nu}{T_\nu}\right) \right] \ .
\end{align}
For initial condition, the above fluctuations do not depend on the direction, thereby characterizing only monopole: $f_\nu = \bar{f}_\nu(1+\Psi_{\nu,0})$.
Then, we obtain
\begin{equation}
\Psi_{\nu,0}(k,q,x_i) \simeq -\dfrac{\text{d}\ln \bar{f}_\nu(q)}{\text{d}\ln q}\left(\dfrac{a\delta\mu_\nu}{q} + \dfrac{q-a\mu_\nu}{q}\dfrac{\delta T_\nu}{T_\nu}\right) \ .
\end{equation}

The initial difference of neutrino density fluctuation is given by
\begin{align}
\delta(\Delta n_
\nu)(x_i) &= \dfrac{g_s}{a^3}\int\dfrac{{\rm d}\bm{q}}{(2\pi)^3}\left[\bar{f}_\nu(q)\Psi_{\nu,0}(k,q,x_i) - \bar{f}_{\bar{\nu}}(q)\Psi_{\bar{\nu},0}(k,q,x_i)\right] \notag \\
&\simeq \dfrac{g_s}{a^3}\int\dfrac{{\rm d}\bm{q}}{(2\pi)^3}\bar{f}_\nu(q)\dfrac{\text{d}\ln \bar{f}_\nu(q)}{\text{d}\ln q}\dfrac{2a\mu_\nu}{q}\left[\left(\dfrac{\delta T_\nu}{T_\nu} -\dfrac{\delta\mu_\nu}{\mu_\nu}\right) + \left(1+2\dfrac{aT_\nu}{q}\dfrac{\text{d}\ln \bar{f}_\nu(q)}{\text{d}\ln q}\right)\dfrac{q}{aT_\nu}\dfrac{\delta T_\nu}{T_\nu} \right] \notag \\
& = \dfrac{g_s}{6}T_\nu^3\dfrac{\mu_\nu}{T_\nu}\left(2\dfrac{\delta T_\nu}{T_\nu} +\dfrac{\delta\mu_\nu}{\mu_\nu}\right) \ .
\label{eq: delta Delta n nu initial}
\end{align}
One can find that this approximation form is also obtained from the expansion of \eqref{eq: BG}.
Now we compute the time evolution of the neutrino density fluctuations.
Similar to Eq.~\eqref{eq: delta Delta n nu initial}, we can expand $\delta(\Delta n_\nu)$ as
\begin{align}
    \delta(\Delta n_
\nu)(x) &= \dfrac{g_s}{a^3}\int\dfrac{{\rm d}\bm{q}}{(2\pi)^3}\left[f_0(q,\mu)\Psi_{\nu,0}(k,q,x) - f_0(q, -\mu)\Psi_{\bar{\nu},0}(k,q,x)\right].
\label{eq: delta Delta nu x 1}
\end{align}
By combining Eqs.~\eqref{eq: solution} and~\eqref{eq: delta Delta nu x 1}, we obtain
\begin{align}
\delta(\Delta n_
\nu)(x) &\simeq 
\left(
\frac{a_i}{a}
\right)^3\delta(\Delta n_\nu)(x_i)j_0(x-x_i) \\
&- \dfrac{g_s}{a^3}\int\dfrac{{\rm d}\bm{q}}{(2\pi)^3}q\left(\dfrac{\text{d}\bar{f}_\nu(q)}{\text{d}q}-\dfrac{\text{d}\bar{f}_{\bar{\nu}}(q)}{\text{d}q}\right)\left[\int_{x_i}^x \left(\psi(x')j_1(x-x') - \dfrac{{\rm d}\phi}{{\rm d}x'}j_0(x-x')\right){\rm d}x' \right] \notag \\
& \simeq 
\left(
\frac{a_i}{a}
\right)^3
\delta(\Delta n_\nu)(x_i)j_0(x-x_i) + \dfrac{g_s}{2}T_\nu^3\dfrac{\mu_\nu}{T_\nu}\left[\int_{x_i}^x \left(\psi(x')j_1(x-x') - \dfrac{{\rm d}\phi}{{\rm d}x'}j_0(x-x')\right){\rm d}x' \right] \ .
\end{align}
Assuming adiabatic condition: $\delta T_\nu/T_\nu = -\psi/2$, we obtain
\begin{align}
\delta(\Delta n_
\nu)(x) &= \Delta\bar{n}_\nu\left[ j_0(x-x_i)\left(-\psi + \dfrac{\delta\mu_\nu}{\mu_\nu}\right) + 3\left[\int_{x_i}^x \left(\psi(x')j_1(x-x') - \dfrac{{\rm d}\phi}{{\rm d}x'}j_0(x-x')\right){\rm d}x' \right]  \right] \ .
\end{align}

During matter domination epoch for CMB epoch, $\psi$ and $\phi$ are almost constant.
Then, 
\begin{align}
\delta(\Delta n_
\nu)(x) &\simeq \Delta\bar{n}_\nu\left[ j_0(x-x_i)\left(-\psi + \dfrac{\delta\mu_\nu}{\mu_\nu}\right) + 3\psi\int_{x_i}^x j_1(x-x'){\rm d}x'  \right] \ .
\end{align}
Using $\int^x_{x_i} j_1(x-x')dx' = 1 - j_0(x-x_i)$, we have
\begin{align}
\delta(\Delta n_
\nu)(x) &\simeq \Delta\bar{n}_\nu\left[ \left(3-4j_0(x-x_i)\right)\psi + j_0(x-x_i)\dfrac{\delta\mu_\nu}{\mu_\nu} \right] \ .
\label{eq: approximate formula for delta Delta n nu}
\end{align}

Now we comment on the relation between $\psi$ and its primordial value $\psi_p$. Regarding the modes that enter the horizon during the radiation-dominated epoch, the amplitude will be suppressed due to the damped oscillations. Also, by taking into account the transition from radiation-dominated era to matter-dominated era, there exists additional suppression factor $\simeq 9/10$. To take it into account, we relate $\psi$ and $\psi_p$ as
\begin{align}
    \psi(x) = \mathcal{T}(k)\psi_p(x),
\end{align}
where $\mathcal{T}(k)$ is a numerically-fitted suppression factor~\cite{Dodelson:2003ft,Inomata:2020lmk}:
\begin{align}
    \mathcal{T}(k)\equiv 
    \frac{9}{10}
    \frac{\ln\left(
    1+0.146x_{\rm eq}
    \right)}{0.146x_{\rm eq}}
    \left(
    1+0.242 x_{\rm eq}
    +
    (1.01 x_{\rm eq})^2
    +
    (0.341 x_{\rm eq})^3
    +
    (0.418 x_{\rm eq})^4
    \right)^{-0.25},
\end{align}
with $x_{\rm eq}\equiv k\tau_{\rm eq}$ with $\tau_{\rm eq}$ being the conformal time at the matter-radiation equality.

The anisotropic cosmic birefringence angle $\delta\beta$ depends on a unit vector $\hat{\bm{n}}$ pointing at the direction where electromagnetic wave comes.
In terms of $\hat{\bm{n}}$, the comoving distance $\bm{x}$ is given by
\begin{equation}
\bm{x} = \Delta\tau\hat{\bm{n}} \ , \qquad \Delta\tau \equiv \tau_0-\tau
\end{equation}
with a displacement of conformal time $\Delta\tau$.
To expand it over the sky, we decompose the anisotropic part into the spherical harmonics components:
\begin{equation}
\delta\beta(\hat{\bm{n}}) = \sum_{\ell m} b_{\ell m}Y_{\ell m}(\hat{\bm{n}}) \ , \qquad \sum_{\ell m} \equiv \sum_{\ell=0}^\infty\sum_{m=-\ell}^{\ell} \ .
\end{equation}
Using this identity
\begin{equation}
\int {\rm d}\hat{\bm{n}} Y_{\ell m}(\hat{\bm{n}})Y^*_{\ell' m'}(\hat{\bm{n}}) = \delta_{\ell\ell'}\delta_{mm'} \ ,
\end{equation}
the harmonic coefficient $b_{lm}$ is given by
\begin{equation}
b_{\ell m} = \int {\rm d}\hat{\bm{n}} Y^*_{\ell m}(\hat{\bm{n}})\delta\beta(\hat{\bm{n}}) = \dfrac{c_\nu}{2M^2}\int {\rm d}\hat{\bm{n}} Y^*_{\ell m}(\hat{\bm{n}})
\int_e^0
{\rm d}t\,
\delta(\Delta n_\nu)(t, \bm{x}) \ , \label{eq: blm11}
\end{equation}
Then, moving to Fourier space 
\begin{equation}
\delta(\Delta n_\nu)(t, \bm{x}) = \int\dfrac{{\rm d}\bm{k}}{(2\pi)^3}\Delta n_{\nu,\bm{k}}(t)e^{i\bm{k}\cdot\Delta\tau\hat{\bm{n}}}
\end{equation}
and using the plane wave expansion
\begin{align}
e^{i\bm{k}\cdot\Delta\tau\hat{\bm{n}}} 
&= 4\pi\sum_{\ell m}i^\ell j_\ell(k\Delta\tau)Y^*_{\ell m}(\hat{\bm{k}})Y_{\ell m}(\hat{\bm{n}}) \ ,
\end{align}
\eqref{eq: blm11} is rewritten as
\begin{equation}
b_{\ell m}
= 
4\pi\dfrac{c_\nu}{2M^2}
\int\dfrac{{\rm d}\bm{k}}{(2\pi)^3}
\int_e^0
{\rm d}t
\Delta n_{\nu,\bm{k}}(t)i^\ell j_\ell(k\Delta\tau)Y^*_{\ell m}(\hat{\bm{k}}) \ . \label{eq: blm22}
\end{equation}

Next, we compute the two-point correlation function of $b_{\ell m}$:
\begin{align}
\langle b_{\ell m}b^*_{\ell'm'} \rangle &= (4\pi)^2\left(\dfrac{c_\nu}{2M^2}\right)^2\int
\dfrac{{\rm d}\bm{k}{\rm d}\bm{k}'}{(2\pi)^6}
\int_e^0
{\rm d}t
\int_e^0
{\rm d}t'
\langle\Delta n_{\nu,\bm{k}}(t)
\Delta n^*_{\nu,-\bm{k}'}(t') \rangle \notag \\
&\times i^\ell (-i)^{\ell'}j_\ell(k\Delta\tau) j_{\ell'}(k'\Delta\tau')Y^*_{\ell m}(\hat{\bm{k}})Y_{\ell' m'}(-\hat{\bm{k}'}) \ .
\end{align}
Then, we can redefine the mode function with the background lepton asymmetry, transfer function and primordial perturbation as
\begin{equation}
\delta(\Delta n_{\nu,\bm{k}})(t) = \Delta \bar{n}_L(t)\left(T_{\psi,k}(t)\psi_{ \bm{k}} + T_{\mu,k}(t)\mu_{p, \bm{k}}\right) \ ,
\end{equation}
where $T_i \ (i = \psi, \mu)$ is a transfer function, where $\mu_{p,\bm{k}}$ is the Fourier component of initial value of $\delta\mu_\nu/\mu_\nu$. The perturbation $\delta\mu_{p,\bm{k}}$ reflects primordial fluctuations
in the lepton asymmetry. We discuss a possible origin of these
fluctuations through the Affleck--Dine mechanism in
Sec.~\ref{sec: AD mechanism as an origin}.
By using Eq.~\eqref{eq: approximate formula for delta Delta n nu}, $T_i(k)$ are approximately given by 
\begin{align}
    T_\mu(k)
    &\simeq j_0(x-x_i) \ ,
    \nonumber \\
    T_\psi(k)
    &\simeq 3-4j_0(x-x_i) \ .
    \label{eq: transfer function for anisotropic CB}
\end{align}
Then, using the definition of power spectrum of perturbation
\begin{equation}
\langle \psi_{p, \bm{k}} \psi^*_{p, -\bm{k}'} \rangle \equiv (2\pi)^3\delta(\bm{k} + \bm{k}')\dfrac{2\pi^2}{k^3}\mathcal{P}_\psi(k) \ , \quad
\langle \mu_{p, \bm{k}} \mu^*_{p, -\bm{k}'} \rangle \equiv (2\pi)^3\delta(\bm{k} + \bm{k}')\dfrac{2\pi^2}{k^3}\mathcal{P}_\mu(k) \ ,
\end{equation}
we obtain
\begin{align}
\langle b_{\ell m}b^*_{\ell'm'} \rangle &= (4\pi)^2\left(\dfrac{\beta}{2M^2}\right)^2\int\dfrac{{\rm d}\bm{k}}{(2\pi)^3}\dfrac{2\pi^2}{k^3}\mathcal{P}_\psi(k)\mathcal{T}(k)^2\int_e^0
{\rm d}t
\int_e^0
{\rm d}t'\,
\bar{n}_L(t)T_{\psi,k}(t)\bar{n}_L(t')T_{\psi,k}(t') \notag \\
&\times i^\ell (-i)^{\ell'}j_\ell(k\Delta\tau) j_{\ell'}(k\Delta\tau')Y^*_{\ell m}(\hat{\bm{k}})Y_{\ell' m'}(\hat{\bm{k}}) + (\psi\rightarrow\mu) \ .
\end{align}
Then, separating ${\rm d}\bm{k} = k^2{\rm d}k {\rm d}\hat{\bm{n}}$ and denoting the angular power spectrum as
\begin{equation}
\langle b_{\ell m}b^*_{\ell'm'} \rangle = C^{\beta\beta}_\ell \delta_{\ell \ell'}\delta_{mm'} \ ,
\end{equation}
we have
\begin{align}
C^{\beta\beta}_\ell &= 4\pi\left(\dfrac{\beta}{2M^2}\right)^2\int\dfrac{{\rm d}k}{k}\mathcal{P}_\psi(k)\left[\int_e^0
{\rm d}t\,
\bar{n}_L(t)T_{\psi,k}(t)j_\ell(k\Delta\tau)\right]^2 + (\psi\rightarrow\mu) \ .
\label{eq: Cl alpha alpha}
\end{align}

\begin{figure}
    \centering
    \includegraphics[width=0.75\linewidth]{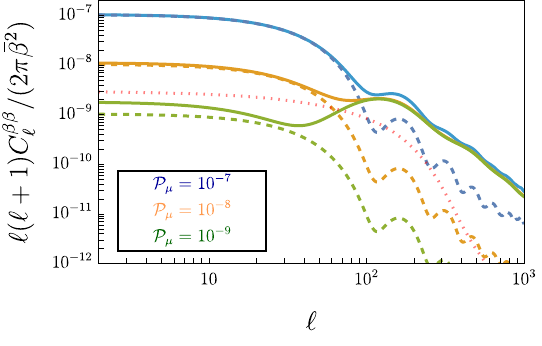}
    \caption{Numerical results of the power spectra of $C_\ell^{\beta\beta}$ for $\mathcal{P}_\mu=10^{-7}$ (blue), $10^{-8}$ (orange) and $10^{-9}$ (green). The solid lines correspond to the numerical results computing Eq.~\eqref{eq: Cl alpha alpha}, while the dashed lines neglects the gravitational effect (only evaluating the second term in Eq.~\eqref{eq: Cl alpha alpha}). Regarding the initial power spectrum $\mathcal{P}_\psi$, we adopt a scale-invariant curvature power spectrum $\mathcal{P}_\zeta = 2.1\times 10^{-9}$ and use the relation $\psi_p=\frac{2}{3}\zeta$ in the superhorizon limit in radiation-dominated era. For reference, the thin-red dotted curve shows the ALP-induced case for $m_\chi=10^{-28.5}\,$eV and $g_{\chi\gamma\gamma}=10^{-18}\,\mathrm{GeV}^{-1}$.}
    \label{fig: anisotropic CB}
\end{figure}

We comute the order of integrations as
\begin{align}
C^{\beta\beta}_\ell = 4\pi\left(\dfrac{\beta}{2M^2}\right)^2\left(\dfrac{g_s}{6}\xi_\nu T_{\nu 0}^3\right)^2\int_e^0
{\rm d}\tau_1a(\tau_1)^{-2}\int_e^0
{\rm d}\tau_2a(\tau_2)^{-2}\int_0^\infty\dfrac{{\rm d}k}{k}\mathcal{P}_\psi(k)\,
T_{k}(\tau_1)j_\ell(k\Delta\tau_1)\,
T_{k}(\tau_2)j_\ell(k\Delta\tau_2) \ ,
\label{eq: C ell beta beta 2}
\end{align}
where $\Delta\tau_i \equiv \tau_0 - \tau_i$.
Regarding the last integration part, assuming the scale invariant primordial spectrum and defining $x_0 \equiv k\tau_0$, we can approximate
\begin{align}
\int_0^\infty\dfrac{{\rm d}k}{k}\mathcal{P}_\psi(k)T_{k}(\tau_1)j_\ell(k\Delta\tau_1)T_{k}(\tau_2)j_\ell(k\Delta\tau_2) \simeq \mathcal{P}_\psi\int_0^\infty\dfrac{{\rm d}x_0}{x_0}T_{k}(\tau_1)j_\ell(k\Delta\tau_1)T_{k}(\tau_2)j_\ell(k\Delta\tau_2)
\end{align}
Due to the dilution of neutrino density by scale factor, we expect that the integration is well approximated during matter domination.
Then, at this time, for time integration we can consider the domain with $k\Delta\tau_i \simeq x_0$.
Regarding the transfer function, it can be written $j_0(x_i-x_{ini}) \simeq j_0(x_i) = j_0(x_0 \tau_i/\tau_0)$.
Then, about an upper limit of integration, we can approximate $x_{max} = 3\ell$ due to the suppression of $j_\ell$.
For smaller $\ell$ and $\tau_i/\tau_0 \ll 1/(3\ell)$, we obtain
\begin{align}
\mathcal{P}_\psi\int_0^\infty\dfrac{{\rm d}x_0}{x_0}T_{k}(\tau_1)j_\ell(k\Delta\tau_1)T_{k}(\tau_2)j_\ell(k\Delta\tau_2) \simeq \mathcal{P}_\psi\int_0^{3\ell}\dfrac{{\rm d}x_0}{x_0}j_\ell(x_0)^2 \sim \dfrac{\mathcal{P}_\psi}{2\ell(\ell+1)}
\end{align}
Then, we can roughly evaluate
\begin{align}
C^{\beta\beta}_\ell \sim \dfrac{2\pi\mathcal{P}_\psi}{\ell(\ell+1)}{\bar\beta}^2 \ . 
\label{eq: analytical approx of anisotropic CB}
\end{align}
In Fig.~\ref{fig: anisotropic CB}, we show the numerical result of the power spectra of $C_l^{\beta\beta}$ for $\mathcal{P}_\mu=10^{-7},10^{-8}, 10^{-9}$.
The solid lines correspond to the numerical results computing Eq.~\eqref{eq: Cl alpha alpha} by assuming the scale invariance of $\mathcal{P}_\mu$, while the dashed lines neglect the gravitational effect (only evaluating the second term in Eq.~\eqref{eq: Cl alpha alpha}). We see that Eq.~\eqref{eq: analytical approx of anisotropic CB} provides a very good approximation for $l\lesssim 30$. However, on small scales with $l\gtrsim 30$, the free streaming of the neutrinos and antineutrinos smooths out the lepton number fluctuations, and therefore the perturbations are suppressed more strongly than predicted by Eq.~\eqref{eq: analytical approx of anisotropic CB}.
Instead, the curvature contribution becomes important on small scales, $\ell \gtrsim 10^2$. 
In Fig.~\ref{fig: anisotropic CB}, for comparison, we also computed the ALP-induced case (thin-red dotted curve) for $m_\chi=10^{-28.5}$\,eV and $g_{\chi\gamma\gamma}=10^{-18}\,{\rm GeV}^{-1}$, where $m_\chi$ is the ALP mass and $g_{\chi\gamma\gamma}$ is the ALP-photon coupling constant, following the procedure in Ref.~\cite{Greco:2022xwj}. On small scales, the suppression is weaker in the neutrino case than in the ALP case. More specifically, while the ALP-induced signal is suppressed approximately as $\ell^{-(5{\text{-}}6)}$, the neutrino-induced signal is suppressed only as $\propto \ell^{-(2{\text{-}}3)}$. The reason can be understood as follows. For larger $\ell$, we can make use of the following Limber approximation~\cite{LoVerde:2008re,Leistedt:2015vwa}:
\begin{align}
   \int_0^\infty \frac{{\rm d}k}{k}
   j_\ell (k\Delta \tau_1)
   j_\ell (k\Delta \tau_2)
   \simeq 
   \frac{\pi\Delta \tau_1}{2(\ell+\frac{1}{2})^3}
   \delta_D(\tau_1-\tau_2),
   \label{eq: Limber approx}
\end{align}
which is valid only for $\ell\gg \Delta\tau_1/|\tau_1-\tau_2|$ (for detail, see Appendix~\ref{append: Limber approximation}). On the other hand, the transfer function~\eqref{eq: transfer function for anisotropic CB} associated with the gravitational contribution approaches a constant on small scales. Using Eq.~\eqref{eq: Limber approx}, and retaining the $k$-dependence of the suppression factor $\mathcal{T}$, we obtain
\begin{align}
\ell(\ell+1)C_\ell^{\beta\beta}
    \propto
    \ell^{-1}
    \mathcal{T}^2
    \left(
    \frac{\ell\tau_{\rm eq}}{\tau_0}
    \right)
\end{align}
with sufficiently large multipoles. For example, at $\ell=500$, the logarithmic slope of $\mathcal{T}$ is approximately
\begin{align}
\left.
\frac{{\rm d}\ln \mathcal{T}(\ell\tau_{\rm eq}/\tau_0)}
     {{\rm d}\ln \ell}
\right|_{\ell=500}
\simeq -0.8,
\end{align}
implying an approximate local scaling of
$\ell(\ell+1)C_\ell^{\beta\beta}\propto\ell^{-2.6}$
near this multipole.

Now we briefly comment on the CMB constraints on lepton number isocurvature perturbations. Lepton number isocurvature perturbations correspond to spatial fluctuations in the neutrino chemical potential, whose impact on the neutrino energy density is up to $\mathcal{O}(\xi_\nu^2)$.
For the small values of $\xi_\nu$ considered in our model, this
suppression leaves the lepton-number isocurvature perturbations
largely unconstrained by current CMB observations.

At the end of this section, we comment on the effect of spatial component of neutrino current $J_i$ on the anisotropic cosmic birefringence. As discussed in detail in Appendix~\ref{Append: effect of spatial current on anisotropic CB}, we confirmed that the effect of $\bm{J}_i$ is subdominant compared to the results presented in this section. The physical reason is that the contribution from the spatial current is primarily sourced by the dipole moment of the neutrino distribution. Since the dipole moment is initially small and develops only at late times, its effect becomes significant only after the background neutrino number density has been substantially diluted as $\Delta n_\nu \propto a^{-3}$. Consequently, the contribution from the spatial current remains subdominant to that from $\delta (\Delta n_{\nu})$.

\section{Lepton asymmetry generation by Affleck-Dine mechanism}
\label{sec: AD mechanism as an origin}
In this section, we consider possible origin of a large lepton asymmetry, which can be the source of the isotropic cosmic birefringence discussed in the previous section. In general, if a large lepton asymmetry is generated in the epoch before the electroweak phase transition, baryon number with a same order is produced via the electroweak sphaleron process~\cite{Kuzmin:1985mm}, which spoils the success of Big Bang nucleosynthesis (BBN). To avoid this problem, one should consider an asymmetry generation well after the electroweak phase transition or some mechanism which suppresses the sphaleron process. The former is possible within so-called $\nu$MSM scenario via decay of light right-handed neutrinos and lepton-to-entropy ratio up to $\sim 7\times 10^{-4}$ can be realized~\cite{Laine:2008pg,Shaposhnikov:2008pf}. On the other hand, in this paper, we consider Affleck-Dine mechanism~\cite{Affleck:1984fy,Dine:1995kz}, which can realize a wider range of lepton asymmetry than $\nu$MSM scenario~\cite{Kawasaki:2002hq,Kawasaki:2022hvx,Kasai:2024diy}. In this model, a CP-violating dynamics of a leptonic complex scalar field, which naturally appears in minimal supersymmetric Standard Model (MSSM), generates a lepton asymmetry. Notably, the above problem is successfully overcome if we consider a formation of classically stable non-topological soliton of the scalar field, known as Q-balls~\cite{Kusenko:1997zq,Enqvist:1997si, Enqvist:2003gh,Kawasaki:2002hq, Kawasaki:2022hvx}. Sphaleron process is largely suppressed inside the Q-ball, since a large VEV of the complex scalar breaks the electroweak symmetry even before the epoch of electroweak phase transition. However, the Q-balls can decay into SM particles such as neutrinos via quantum effect, which typically takes place well after the electroweak phase transition, enabling us to successfully explain the origin of a lepton asymmetric SM plasma before BBN. 

Thus, in this chapter, we consider the AD mechanism as a possible origin for the asymmetric neutrino current which induces the isotropic cosmic birefringence. 
\subsection{Coherent generation of lepton asymmetry}
Here, we discuss the Affleck-Dine mechanism as a possible origin for the cosmic birefringence, induced by the coupling given by Eq.~\eqref{eq : vector CS coupling}. Now we briefly review the Affleck-Dine leptogenesis scenario.

AD leptogenesis utilizes flat directions in the scalar potential of the minimal supersymmetric standard model (MSSM). 
Here, we consider one of the leptonic flat directions and call it the AD field.
The potential of the AD field is lifted by SUSY breaking effects and the existence of a cutoff at the Planck scale, $M_\text{Pl}$.

During inflation, the AD field has a potential given by~\cite{Dine:1995kz}
\begin{align}
    V(\phi) 
    & =
    m_\phi^2|\phi|^2 - cH^2|\phi|^2
    + V_\text{NR} + V_\text{A}
    \nonumber\\
    & =
    m_\phi^2|\phi|^2 - cH^2|\phi|^2
    + |\lambda|^2 \frac{|\phi|^{2(n-1)}}{M_{\rm{Pl}}^{2(n-3)}} 
    + \left(\lambda a_M \frac{m_{3/2}\phi^n}{n M_{\rm{Pl}}^{n-3}}
    + \mathrm{h.c.}\right)
    \ ,
    \label{AD field potential}
\end{align}
where $m_\phi$ is the soft SUSY breaking mass of the AD field, $m_{3/2}$ is the gravitino mass, and $c$, $a_M$, and $\lambda$ are dimensionless constants. 
Here, we take $m_\phi\simeq m_{3/2}$, which holds for sufficiently large field values.
The third term $V_\text{NR}$ is called the non-renormalizable term, and the fourth term $V_\text{A}$ is the A-term, which violates a global $U(1)$ symmetry associated with lepton number.
$n\, (\ge 4)$ is an integer depending on the choice of the flat direction and non-renormalizable superpotential.
As a concrete flat direction~\cite{Gherghetta:1995dv}, we consider $\mu^{c}l_{L_e}l_{L_\tau}$ or $\tau^{c}l_{L_e}l_{L_\mu}$ direction,
where $a^c$ is the charge conjugation of an $SU(2)_{\rm{W}}$ singlet charged lepton with flavor $a =e, \mu, \tau$, and
$l_{L_a}$ is an $SU(2)_{\rm{W}}$ doublet lepton. In this case, the flat direction is lifted by the superpotentials of the form $W\propto (LLe)^2$, and thus we take $n=6$. We neglect the thermal potential of the AD field induced by interactions with inflaton decay products, which is justified with the sufficiently low reheating temperature, $T_{\rm R} \lesssim 10^5\,\mathrm{GeV}$, assumed in this work.

After inflation, $H$ decreases and becomes smaller than the SUSY breaking mass at some point.
While $H\gtrsim m_{\phi}$, the AD field has a non-zero field value given by
\begin{equation}
    |\phi|
    \simeq
    \varphi_{\rm inf}
    \equiv
    \left(
        \frac{\sqrt{c}HM_{\rm{Pl}}^{n-3}}{|\lambda|}
    \right)^{\frac{1}{n-2}}
    \ ,
    \label{eq : radial profile of AD field right after inf}
\end{equation}
due to the balance between the Hubble-induced mass and the non-renormalizable term.
When $H$ becomes comparable to $m_{\phi}/\sqrt{c}$, the origin of the field space is stabilized due to the soft mass term, and when $H\sim m_\phi$, the AD field starts to oscillate around the origin. 
At this period, the field value is given by 
\begin{equation}
    |\phi|
    \simeq
    \varphi_{\rm{osc}}
    \equiv 
    \left(
        \frac{
        m_{\phi} M_{\rm{Pl}}^{n-3}}{|\lambda|}
    \right)^{\frac{1}{n-2}}
    \ .
    \label{eq: varphi osc}
\end{equation}
At the same time, the A-term kicks the AD field in the phase direction.
Consequently, the AD field acquires a lepton number proportional to $\varphi^2 \dot{\theta}_\mathrm{AD}$, where $\varphi$ and $\theta_\mathrm{AD}$ are the radial and phase components of $\phi$, i.e., $\phi \equiv \varphi e^{i\theta_\mathrm{AD}}$. 
The resulting lepton number density
is written as
\begin{equation}
    n_L(t_\mathrm{osc}) \simeq \epsilon\, m_\phi \varphi_\mathrm{osc}^2 
    \ ,
    \label{resulting lepton number in AD leptogenesis}
\end{equation}
where $\epsilon$ is the efficiency parameter depending on $n$, the initial phase $\theta_i$ of the AD field, and the phases of $a_M$ and $\lambda$. The concrete relation is given by
\begin{equation}
    \epsilon = 
    \frac{4|a_M|}
    {3\left(1+\frac{n-4}{n-2}\right)}
    \sin
    \left(
    \arg
    \left(
    \lambda a_M
    \right)+n\theta_i
    \right),
\end{equation}
where we have substituted $n=6$ in the last equality.

From now on, we consider the gauge-mediated SUSY breaking scenario.
After the AD field starts to oscillate, the AD field potential is given by
\begin{align}
    V
    &=
    V_{\rm{gauge}} + V_{\rm{gravity}}
    \ .
    \label{gauge mediation and gravity mediation potential}
\end{align}
Here, $V_{\rm{gauge}}$ is the contribution from gauge-mediation given by~\cite{deGouvea:1997afu,Kusenko:1997si}
\begin{align}
    V_{\rm{gauge}}
    =
    M_{\rm{F}}^4 \left( \ln \frac{|\phi|^2}{M_S^2} \right)^2
    \ ,
    \label{eq:Vgauge}
\end{align}
where $M_{\rm{F}}$ is the SUSY breaking scale, and $M_S\sim M_F^2/M_s$, with $M_s$ being the mass scale of gauginos, is the scale of the messenger sector.
Here, we consider the filed regime with $|\phi|\gg M_S$.
$V_{\rm{gravity}}$ is the contribution from gravity-mediation given by
\begin{align}
    V_{\rm{gravity}}
    =
    m_{3/2}^2
    \left[
        1 + K\ln\left( \frac{|\phi|^2}{M_{\rm{Pl}}^2} \right)
    \right]
    |\phi|^2
    \ ,
    \label{eq:Vgravity}
\end{align}
where $K$ is a dimensionless constant, depending on the coupling with other fields. In the following, we consider the both cases, $K<0$ and $K>0$. In the following, we define the critical value of AD field $\varphi_{\rm{eq}}\equiv M_{\rm{F}}^2/m_{3/2}$ for convenience, such that $V_{\rm{gravity}}$ is dominant in the case with $\varphi\gtrsim\varphi_{\rm{eq}}$ while $V_{\rm{gauge}}$ is dominant with $\varphi\lesssim\varphi_{\rm{eq}}$. 

\subsection{Formation and decay of Q-balls}
\subsubsection{The case with $K>0$}
Assuming $\varphi_{\rm{osc}}\gg\varphi_{\rm{eq}}$, $V_{\rm{gravity}}$ is dominant at the onset of AD field oscillations. In this period, the coherent rotation of AD field is stable. On the other hand, once the $V_{\rm{gauge}}$ becomes dominant, i.e. $\varphi\lesssim \varphi_{\rm{eq}}$, the coherent field rotation is no longer stable, and it forms spherical clumps called Q-balls. Then, the initial charge $Q_{\rm D}^{\rm init}$, mass $M_{\rm D}$, radius $R_{\rm D}$, and energy per charge $\omega_{\rm D}$ of a delayed-type Q-ball are given by~\cite{Kasuya:1999wu,Kasuya:2001hg,Hisano:2001dr}
\begin{align}
    |Q_{\rm{D}}^{\mathrm{init}}|
    & \simeq
    \gamma_{\rm{D}} \frac{\varphi_{\rm{eq}}^4}{M_{\rm{F}}^4}
    \simeq
    \gamma_{\rm{D}} 
    \left(
        \frac{M_{\rm{F}}}{m_{3/2}}
    \right)^4
    \ , 
    \label{Q_G}
    \\[0.4em]
    M_{\rm D} & \simeq \frac{4\sqrt{2}\pi}{3}\zeta M_{\rm{F}} 
    |Q_{\rm{D}}|^{\frac{3}{4}}
    \ ,
    \label{eq : Q-ball property1}
    \\[0.4em]
    R_{\rm D} & \simeq \frac{1}{\sqrt{2}\zeta}M_{\rm{F}}^{-1}
    |Q_{\rm{D}}|^{\frac{1}{4}}
    \ ,
    \label{eq : Q-ball property2}
    \\[0.4em]
    \omega_{\rm D} & \simeq \sqrt{2}\pi \zeta M_{\rm{F}}
    |Q_{\rm{D}}|^{-\frac{1}{4}}
    \ ,
    \label{eq : Q-ball property3}
\end{align}
where $\gamma_{\rm{D}}\simeq 6\times10^{-4}$ in the case with $|\epsilon| \simeq 1$, which we consider in this paper,
and $\zeta \simeq 2^{1/4}(c_0/\pi)^{1/2}$ is a numerical factor where $c_0\equiv 4.8\ln (m_\phi/\sqrt{2}\omega_{\rm D})+7.4$. In the following, we adopt $\zeta=3.5$, which is obtained by substituting $m_{3/2}=1$\,GeV, $M_{\rm{F}}=10^7$\,GeV and $m_\phi=10^4$\,GeV. 
As mentioned before, the AD field is composed of sleptons in our setup. In this case, the main decay channel of Q-balls is via the two-body annihilation of AD field into SM neutrinos via exchange of Zino/Higgsino.~\footnote{The mass of AD field inside the Q-ball, $\omega_{\rm D}$, is the order of gravitino mass (see Eq.~\eqref{eq : Q-ball property3}), which is typically much smaller than the masses of gaugino/higgsino in the gauge-mediation models. Thus, the decay of a single AD field into a gaugino/higgsino and a neutrino is kinematically forbidden. Instead, two-body annihilation becomes a dominant decay channel.} 
The decay rate of Q-ball per unit charge, denoted as $\Gamma_{Q}$, is bounded by the Pauli's exclusion principle of outgoing neutrinos~\cite{Cohen:1986ct}. 
The upper bound, denoted as $\Gamma_Q^{\rm sat}$, is estimated as~\cite{Cohen:1986ct,Kawasaki:2012gk}
\begin{equation}
    \Gamma_{Q}^{\rm sat}
    \equiv 
    \frac{N_l}{|Q_\mathrm{D}|}\frac{\omega_{\rm D}^3}{12\pi^2}4\pi R_{\rm D}^2
    \ ,
    \label{Qball decay rate}
\end{equation}
where $N_l=3$ is the number of SM neutrino species.~\footnote{We can safely neglect the decay of Q-balls into gravitinos, since the effective coupling of AD field with gravitino is suppressed by $M_{\rm Pl}$~\cite{Kasuya:2012mh}.} According to the detailed numerical calculation~\cite{Multamaki:1999an,Kawasaki:2012gk}, a saturated decay takes place when $\varphi_Q^2\gg M_{s}\omega_{\rm D}$ with $\varphi_Q$ being the VEV of $\varphi$ inside the Q-ball, which always holds in the parameter region of our interest. Thus, we set $\Gamma_Q=\Gamma_Q^{\rm sat}$ throughout this work.

Now we estimate the lepton-to-entropy ratio from the decay of Q-balls. Due to the charge neutrality, the resultant lepton asymmetry from the decay of Q-balls is distributed only in the neutrino sector.~\footnote{Depending on the type of the AD field, a lepton asymmetry may be partially distributed into the charged-lepton sector. Even in such cases, the total charge neutrality can be maintained when the asymmetries are summed over all flavors.
However, in the typical parameter region considered in this work, the cosmic temperature immediately after the decay of Q-balls is sufficiently lower than the masses of charged leptons other than the electron. As a consequence, imposing the charge neutrality condition renders the asymmetry in the charged sector negligibly small.}

The resulting lepton-to-entropy ratio is given by

\begin{align}
    L_{e} \equiv \dfrac{n_L(t_{\rm dec})}{s(t_{\rm dec})} &\simeq \frac{n_L(t_{\rm{osc}})}{m_{3/2}^2\varphi_{\rm{osc}}^2} 
    \frac{\rho_{\rm{Q}}(t_{\rm{dec}})}{s(t_{\rm{dec}})}
    \simeq 
    \frac{3}{4}
    \frac{T_{\rm{dec}}n_L(t_{\rm{osc}})}{m_{3/2}^2\varphi_{\rm{osc}}^2}
    \frac{\rho_Q(t_{\rm{dec}})}{\rho_{\rm{tot}}(t_{\rm{dec}})},
    \label{Relation between etaL gravitino mass and Tdecay}
\end{align}
with $\rho_Q$ being the energy density of Q-balls.

To estimate the $\rho_{\rm{Q}}/\rho_{{\rm{tot}}}$ at $t=t_{\rm{dec}}$, we define the energy fraction of the Affleck-Dine field before reheating as $f_{{\rm{osc}}}$, which is approximately given by

\begin{align}
    f_{{\rm{osc}}}\simeq 
    \frac{m_{\phi}^2\varphi_{\rm{osc}}^2}{3M_{\rm{Pl}}^2H_{\rm{osc}}^2}
    \simeq 
    \frac{1}{3}
    \left(
    \frac{\varphi_{\rm{osc}}}{M_{\rm{Pl}}}
    \right)^2,
\end{align}
where we have used $m_{\phi}\simeq H_{\rm{osc}}$. Here, we have approximated Eq.~\eqref{AD field potential} by the first term, which is valid after the onset of the coherent AD field oscillation. This energy fraction is approximately conserved before reheating. After the reheating, the Hubble parameter is given by
\begin{align}
    H
    =
    H_{\rm{R}}
    \sqrt{
    \frac{
    \left(
    \frac{a_{\rm{R}}}{a}
    \right)^4
    +f_{\rm{osc}}
    \left(
    \frac{a_{\rm{R}}}{a}
    \right)^3
    }
    {1+f_{\rm{osc}}}
    },
\end{align}
where $H_{\rm{R}}$ and $a_{\rm{R}}$ denote the Hubble parameter and scale factor at reheating. The relation between $H_{\rm{R}}$ and $T_{\rm{R}}$ is given by
\begin{align}
    H_{\rm{R}} = \sqrt{\frac{\pi^2}{90}g_*(T_{\rm{R}})}
    \frac{T_{\rm{R}}^2}{M_{\rm{Pl}}},
\end{align}
where we have neglected the energy fraction of the AD field at the reheating. 

Since the Q-balls decay when $H=\Gamma_Q$, the $\rho_Q(t_{\rm{dec}})/\rho_{\rm{tot}}(t_{\rm{dec}})$ is then given by
\begin{align}
    \frac{\rho_Q(t_{\rm{dec}})}{\rho_{\rm{tot}}(t_{\rm{dec}})}
    \simeq
    \frac{f_{\rm{osc}}}{f_{\rm{osc}}+\frac{a_{\rm{R}}}{a_{\rm{dec}}}},
\end{align}
where $a_{\rm{R}}/a_{\rm{dec}}$ satisfies
\begin{align}
    \left(
    \frac{a_{\rm{R}}}{a_{\rm{dec}}}
    \right)^4
    +f_{\rm{osc}}
    \left(
    \frac{a_{\rm{R}}}{a_{\rm{dec}}}
    \right)^3
    =
    (1+f_{\rm{osc}})
    \left(
    \frac{\Gamma_Q}{H_{\rm{R}}}
    \right)^2,
    \label{eq : aR over adec}
\end{align}
and unless $a_{\rm{R}} \simeq f_{\rm osc}a_{\rm{dec}}$, then approximately given by
\begin{align}
    \frac{a_{\rm{R}}}{a_{\rm{dec}}}
    \simeq
    {\rm Max}
    \left[
    \left(
    \frac{1+f_{\rm osc}}{f_{\rm osc}}
    \right)^{1/3}
    \left(
    \frac{\Gamma_Q}{H_{\rm R}}
    \right)^{2/3},\;
    \left(
    1+f_{\rm osc}
    \right)^{1/4}
    \sqrt{\frac{\Gamma_Q}{H_{\rm R}}}
    \right].
\end{align}

Q-balls decay almost instantaneously when the decay rate $\Gamma_Q$ becomes equal to the Hubble parameter. 
Thus, $T_\mathrm{dec}$ is given by
\begin{align}
    T
    \simeq
    T_{\rm{dec}} 
    &\equiv 
    \left(\frac{90}{\pi^2 g_{\ast}(T_{\rm{dec}})}\right)^{1/4}\sqrt{M_{\rm{Pl}}\Gamma_{Q}}\nonumber
    \\
    &\simeq
    11\,{\rm{MeV}}\times \left(\frac{g_{\ast}(T_{\rm dec})}{10.75}\right)^{-1/4}
    \left(\frac{m_{3/2}}{1\,{\rm{GeV}}}\right)^{5/2}\left(\frac{M_{\rm{F}}}{10^7\,{\rm{GeV}}}\right)^{-2}
    \ .
    \label{Tdecay}
\end{align}
Next, the resultant lepton-to-entropy ratio is given by

\begin{align}
    L_{e} &\simeq \frac{3}{4}
\frac{n_L(t_{\rm{osc}})}{m_{3/2}^2\varphi_{\rm{osc}}^2} 
    \frac{\rho_{\rm{Q}}(t_{\rm{dec}})}{s(t_{\rm{dec}})}
    \simeq 
    \frac{3}{4}\epsilon\frac{T_{\rm{dec}}}{m_{3/2}}
    \frac{f_{\rm{osc}}}{f_{\rm{osc}}+\frac{a_{\rm{R}}}{a_{\rm{dec}}}},
    \label{eq: L init}
    \nonumber \\
    &\simeq
    8.1\times 10^{-3}\,
    \epsilon
    \times
    \begin{cases}
    \displaystyle
    \left(
    \frac{m_{3/2}}{1\,{\rm GeV}}
    \right)^{3/2}
    \left(
    \frac{M_{\rm F}}{10^7\,{\rm GeV}}
    \right)^{-2} &(\varphi_{\rm osc}\gtrsim \varphi_{\rm cr,D})
    \\[0.8em]
    \displaystyle
    \left(
    \frac{m_{3/2}}{1\,{\rm GeV}}
    \right)^{-1}
    \left(
    \frac{T_{\rm R}}{10^5\,{\rm GeV}}
    \right)
    \left(
    \frac{\varphi_{\rm osc}}{\varphi_{\rm cr,D}}
    \right)^2
    &(\varphi_{\rm osc}\lesssim \varphi_{\rm cr,D})
    \end{cases}
    ,
\end{align}
where $\varphi_{\rm cr,D}$ is defined as
\begin{align}
    \varphi_{\rm cr,D}
    \equiv
    1.4\times 10^{15}\,{\rm GeV}
    \times 
    \left(
    \frac{g_{*,s}(T_{\rm dec})}{10.75}
    \right)^{-1/8}
    \left(
    \frac{m_{3/2}}{1\,{\rm GeV}}
    \right)^{5/4}
    \left(
    \frac{M_{\rm F}}{10^7\,{\rm GeV}}
    \right)^{-1}
    \left(
    \frac{T_{\rm R}}{10^5\,{\rm GeV}}
    \right)^{-1/2}.
\end{align}

Next, we consider the baryon asymmetry produced from the evaporation process of Q-balls until the sphaleron process becomes out of equilibrium. Q-balls continuously lose and emit their charge due to the interactions with neutrinos and gauginos in the thermal background.
This process is called evaporation and its rate is written as
\begin{align}
    \Gamma_\mathrm{evap}
    \equiv
    \frac{\mathrm{d}Q}{\mathrm{d}t}\bigg|_\mathrm{evap}
    \sim
    -4\pi
    \tilde{\chi}\left(\mu_Q-\mu_\mathrm{plasma}\right)T^2R_{\rm D}^2
    \simeq -4\pi\tilde{\chi}\omega_{\rm D}T^2 R_{\rm D}^2
    \ ,
    \label{eq: evap rate}
\end{align}
where $\mu_{Q,\,(\mathrm{plasma})}$ is the chemical potential for lepton number inside (outside) the Q-ball.
Here we use $\mu_Q =\omega_Q$ and $\mu_\mathrm{plasma}\ll \mu_Q$, and the $\tilde{\chi}$ depends on cosmic temperature given by~\cite{Laine:1998rg}
\begin{align}
    \tilde{\chi}(T)
    =
    \begin{cases}
        1 & (T>M_s)\\
        \left(\frac{T}{M_s}\right)^2 & (T<M_s)
    \end{cases}
    \ .
    \label{eq: rate coefficient}
\end{align}
where $M_s$ is the typical mass scale of the gauginos involved in the evaporation process. 

For efficient Q-ball evaporation, the evaporated charge should be diluted away from the Q-ball surface by diffusion~\cite{Banerjee:2000mb}.
The diffusion rate is given by
\begin{align}
    \Gamma_\mathrm{diff}
    \equiv
    \frac{\mathrm{d}Q}{\mathrm{d}t}\bigg|_\mathrm{diff}
    \sim -4\pi DR_{\rm D}\mu_QT^2
    \simeq -4\pi DR_{\rm D}\omega_{\rm D}T^2
    \ ,
    \label{eq: diff rate}
\end{align}
where the diffusion constant $D\equiv A/T$ is estimated with a numerical constant $A=4\,\text{--}\,6$. Here, we take $A=5$ as a reference value.
When the diffusion rate $\Gamma_\mathrm{diff}$ is smaller than $\Gamma_\mathrm{evap}$, the evaporation process is controlled by diffusion.
Thus, the rate of the evaporation process is determined by the smaller one of $\Gamma_\mathrm{evap}$ and $\Gamma_\mathrm{diff}$, and is given by $\Gamma_\mathrm{diff}$ at high temperatures and $\Gamma_\mathrm{evap}$ at low temperatures.
For later convenience, we define the temperature $T_{\rm eq}$ for $\Gamma_\mathrm{diff}(T_{\rm eq})=\Gamma_\mathrm{evap}(T_{\rm eq})$.

By integrating the above evaporation rates, we obtain total evaporated baryon charge $\Delta Q_b$ as ~\cite{Kasuya:2014ofa,Kasai:2024diy}\footnote{
Here, we assume $T_{\rm{ew}}<T_{\rm{eq}}$.
}

\begin{align}
   \frac{\Delta Q_b}{Q_{\rm D}^\mathrm{init}}
    \simeq 
    \begin{cases}
    -
    \frac{16\pi}{23\sqrt{5}}
    \frac{ \sqrt{g_*(T_{\rm{R}})} }{ g_*(T_{\rm{ew}}) }
    \left(
        32 + 96 \log \frac{T_{\rm{eq}}}{T_{\rm{ew}}}
    \right)
    \frac{M_{\rm{Pl}}T_{\rm{R}}^2}{\zeta M_s^2M_F}
    |Q_{\rm D}^\mathrm{init}|^{-3/4} 
    &
    (T_{\rm{R}}<T_{\rm{ew}} < T_\mathrm{eq})
    \vspace{1mm} \\ 
    -\frac{16\pi }{23\sqrt{5}}\frac{1}{\sqrt{g_*(T_{\rm{ew}})}}
    \left(
        47
        - 15\frac{T_\mathrm{ew}^2}{T_\mathrm{R}^2}
        +96\log\frac{T_{\rm{eq}}}{T_\mathrm{R}}
    \right)
    \frac{M_\mathrm{Pl} T_\mathrm{R}^2}
    {\zeta M_s^2M_{\rm{F}}}
    |Q_{\rm D}^\mathrm{init}|^{-3/4}
    &
    (T_\mathrm{ew} < T_\mathrm{R} < T_{\rm{eq}})
    \vspace{1mm} \\
    -\frac{16\pi }{23\sqrt{5}}\frac{1}{\sqrt{g_*(T_{\rm{ew}})}}
    \left(
        45 - 15\frac{T_\mathrm{ew}^2}{T_{\rm{eq}}^2}
        + 2 \frac{T_\mathrm{eq}}{T_\mathrm{R}}
    \right)
    \frac{M_\mathrm{Pl} T_{\rm{eq}}^2}{\zeta M_s^2M_{\rm{F}}}
    |Q_{\rm D}^\mathrm{init}|^{-3/4}
    &
    (T_\mathrm{ew} < T_{\rm{eq}} < T_\mathrm{R})
    \end{cases}
    \ .
    \label{The total evaporation charge}
\end{align}
Here, we have used the relation between the physical time and the cosmic temperature as
\begin{align}
    \frac{{\rm d}t}{{\rm d}T}
    =
    \begin{cases}
        \frac{288}{\pi^2 g_{*}}
        M_{\rm Pl}^2H(T_{\rm R})
        T^{-5} &
        (T\gtrsim T_{\rm R})
        \vspace{1mm} 
        \\
        \frac{3}{\pi}
        \sqrt{\frac{10}{g_{*}}}
        M_{\rm Pl}T^{-3}
        &
        (T\lesssim T_{\rm R})
    \end{cases}
    \,.
\end{align}

Here, $T_{\rm{eq}}\sim (M_s^2M_{\rm{F}})^{1/3}
|Q_G^\mathrm{init}|^{-1/12}$
is the temperature at which the evaporation rate of Q-balls and the diffusion rate of leptons evaporated from Q-balls becomes equal, and $T_{\rm{ew}}$ is the temperature at electroweak phase transition. We assumed $g_*(T)=g_*(T_{\rm{ew}})$ for $T>T_{\rm{ew}}$.
Then, if $T_\mathrm{R} < T_\mathrm{ew} < T_\mathrm{eq}$, the total baryon asymmetry induced from the Q-ball evaporation process can be evaluated as
\begin{align}
    \eta_b &\simeq L_e\left(\frac{\Delta Q_b}{Q_{\rm{D}}^{\rm init}}\right) \ .
    \label{Etab}
\end{align}
We require that this resulting baryon asymmetry does not spoil the BBN, that is, $|\eta_b|<|\eta_b^{\rm{obs}}|\sim 10^{-10}$, which leads to the upper bound of $T_{\rm{R}}$ in terms of $M_s,m_{3/2}$, and $M_{\rm{F}}$.
Finally, we consider the gravitino problem. The gravitino density parameter $\Omega_{3/2}$ is approximately given by~\cite{Kawasaki:2017bqm}
\begin{align}
    \Omega_{3/2}h^2
    &\simeq 
    0.71\left(\frac{m_{3/2}}{0.5\,{\rm{GeV}}}\right)^{-1}\left(\frac{M_{\tilde{g}}}{10^4\,{\rm{GeV}}}\right)^2\left(\frac{T_{\rm{R}}}{10^5\,{\rm{GeV}}}\right)
    r_s^{-1}
    \ ,
    \label{eq : gravitino density parameter}
\end{align} 
where $M_{\tilde{g}}$ is the gluino mass. 
Here, we added a factor $r_s$, which represents the dilution of gravitinos due to entropy production via Q-balls decay and is defined by
\begin{equation}
r_s \equiv \dfrac{s_{\rm dec,after}}{s_{\rm dec,before}}
    =
    \dfrac{T_{\rm dec,before}}{T_{\rm dec,after}}\left(
    1+
    f_{\rm{osc}}
    \frac{a_{\rm{dec}}}{a_{\rm{R}}}
    \right)
    \ ,
\end{equation}
where $T_{\rm dec,before}$ ($T_{\rm dec,after}$) is the background temperature right before (after) the decay of Q-balls, and $s_{\rm dec,before}$ ($s_{\rm dec,after}$) is the total entropy density right before (after) the decay of Q-balls. 
Since the energy density of radiation from the inflaton decay scales as $\propto a^{-4}$ while that of Q-balls scales as $\propto a^{-3}$, the energy fraction of Q-balls compared to the radiation right before the decay is given by $1+f_{\rm osc}a_{\rm dec}/a_{\rm R}$. By assuming the instantaneous decay of Q-balls and using the energy conservation before and after the decay, we obtain
\begin{align}
    \frac{\pi^2}{30}
    g_{*}(T_{\rm dec,after})
    T_{\rm dec,after}^4
    &=
    \frac{\pi^2}{30}
    g_{*}(T_{\rm dec,before})
    T_{\rm dec,before}^4
    +
    \rho_{\rm Q}(t_{\rm dec})
    \nonumber \\
    &=
    \frac{\pi^2}{30}
    g_{*}(T_{\rm dec,before})
    T_{\rm dec,before}^4
    \left(
    1+
    f_{\rm{osc}}
    \frac{a_{\rm{dec}}}{a_{\rm{R}}}
    \right).
\end{align}
Thus, we can relate $T_{\rm dec,before}$ and $T_{\rm dec, after}$ as
\begin{align}
    \frac{T_{\rm dec,before}}{T_{\rm dec,after}}
    =
    \left[
    \frac{g_*(T_{\rm dec,before})}{g_*(T_{\rm dec,after})}
    \left(
    1+f_{\rm osc}
    \frac{a_{\rm dec}}{a_{\rm R}}
    \right)
    \right]^{-1/4}
    \simeq
    \left(
    1+f_{\rm osc}
    \frac{a_{\rm dec}}{a_{\rm R}}
    \right)^{-1/4}
\end{align}
and we obtain
\begin{align}
    r_s
    =
    \left(
    1+
    f_{\rm{osc}}
    \frac{a_{\rm{dec}}}{a_{\rm{R}}}
    \right)^{3/4}
    \ .
    \label{eq : definition of sQ}
\end{align}

\begin{figure}
    \centering
    \includegraphics[width=0.8\linewidth]{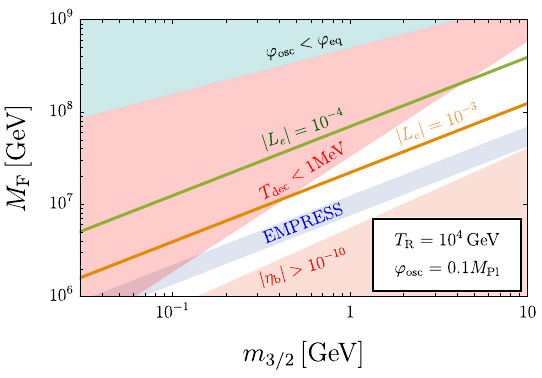}
    \caption{$m_{3/2}$-$M_{\rm F}$ contour to explain the origin of lepton asymmetry by the decay of delayed-type Q-balls. The thin blue region can realize the range of electron-neutrino asymmetry parameter $\xi_{\nu_e}=0.05^{+0.03}_{-0.02}$, which is suggested by the EMPRESS data~\cite{Matsumoto:2022tlr}. The red region is not allowed because the cosmic temperature immediately after the decay of Q-balls is below 1 MeV, which is insufficient to thermalize the neutrinos produced by the decay of Q-balls. The thin red region is not allowed because of the baryon number overproduction. We also note that the light gray region in the upper part indicates the parameter space where our assumption, $\varphi_{\rm osc} > \varphi_{\rm eq}$, is no longer valid.}
    \label{fig : delayed-type case}
\end{figure}
In Fig.~\ref{fig : delayed-type case}, we show the $m_{3/2}$-$M_{\rm F}$ contour to explain the origin of lepton asymmetry.
One can find that there exists the allowed parameter region to explain the isotropic cosmic birefringence with a large lepton asymmetry favored by the recent observation of Subaru EMPRESS and CMB ACT observatory without overproduction of baryon asymmetry, nor spoiling the success of BBN.

\subsubsection{The case with $K<0$}
Here, we consider the case with $K<0$. In this case, the AD field has a spatial instability at the onset of coherent oscillation, leading to Q-ball formation. This type of Q-balls are called New-type Q-balls. The initial charge $Q_{\rm N}^{\rm init}$, mass $m_{\rm N}$, radius $R_{\rm N}$, and energy density per unit charge $\omega_{\rm N}$ of these Q-balls are given by~\cite{Hisano:2001dr, Kasuya:2014ofa}
\begin{align}
    |Q_{\rm{N}}^{\mathrm{init}}|
    & \simeq
    \gamma_{\rm N}
    \left(
    \frac{\varphi_{\rm osc}}{m_{3/2}}
    \right)^2
    \ , 
    \label{Q_N}
    \\[0.4em]
    R_{\rm N} &\simeq 
    |K|^{-1/2}m_{3/2}^{-1}
    \ ,
    \label{eq :  New type Q-ball prop 2}
    \\[0.4em]
    \omega_{\rm N}
    &\simeq
    m_{3/2}
    \ ,
    \label{eq :  New type Q-ball prop 4}
    \end{align}
where $\gamma_{\rm N}\simeq 0.02$ is a numerical coefficient~\cite{Kasuya:2001hg,Kasuya:2014ofa}.
By the replacement of $Q_{\rm D}\to Q_{\rm N}$, $R_{\rm D}\to R_{\rm N}$ and $\omega_{\rm D}\to\omega_{\rm N}$ in Eq.~\eqref{Qball decay rate},~\eqref{eq: evap rate} and~\eqref{eq: diff rate}, respectively, we obtain the decay rate and the evaporation rate of Q-balls as
\begin{align}
    \Gamma_Q
    &\simeq 
    \frac{N_l}{|Q_\mathrm{N}|}\frac{\omega_{\rm N}^3}{12\pi^2}4\pi R_{\rm N}^2,
    \label{eq: Gamma Q in New Type}
    \\[0.4em]
    \Gamma_{\rm evap}
    &\simeq
    -4\pi\tilde{\chi}\omega_{\rm N}T^2 R_{\rm N}^2,
    \label{eq: Gamma evap in New Type}
    \\[0.4em]
    \Gamma_{\rm diff}
    &\simeq
   -4\pi DR_{\rm N}\omega_{\rm N}T^2.
   \label{eq: Gamma diff in New Type}
\end{align}

By equating the Hubble parameter to Eq.~\eqref{eq: Gamma Q in New Type} with Eqs.~\eqref{Q_N},~\eqref{eq :  New type Q-ball prop 2} and~\eqref{eq :  New type Q-ball prop 4}, we obtain the cosmic temperature at the decay of Q-balls, $T_\mathrm{dec}$, as
\begin{align}
    T
    \simeq
    T_{\rm{dec}} 
    &\equiv 
    \left(\frac{90}{\pi^2 g_{\ast}(T_{\rm{dec}})}\right)^{1/4}\sqrt{M_{\rm{Pl}}\Gamma_{Q}}\nonumber
    \\
    &\simeq
    6.0\,{\rm MeV}
    \times
    \left(
    \frac{g_{*,s}}{10.75}
    \right)^{-1/4}
    \left(
    \frac{m_{3/2}}{10\,{\rm GeV}}
    \right)^{3/2}
    \left(
    \frac{\varphi_{\rm osc}}{10^{14}\,{\rm GeV}}
    \right)^{-1}
    \ .
    \label{Tdecay in New Type}
\end{align}

The resultant lepton-to-entropy ratio after the decay of Q-balls, $L_{e}$, is given as
\begin{align}
    L_{e} &\simeq \frac{3}{4}
\frac{n_L(t_{\rm{osc}})}{m_{3/2}^2\varphi_{\rm{osc}}^2} 
    \frac{\rho_{\rm{Q}}(t_{\rm{dec}})}{s(t_{\rm{dec}})}
    \simeq 
    \frac{3}{4}\epsilon\frac{T_{\rm{dec}}}{m_{3/2}}
    \frac{f_{\rm{osc}}}{f_{\rm{osc}}+\frac{a_{\rm{R}}}{a_{\rm{dec}}}}
    \nonumber \\[0.6em]
    &\simeq
    4.5\times 10^{-4}
    \epsilon\times
    \begin{cases}
    \displaystyle
    \left(
    \frac{g_{*,s}}{10.75}
    \right)^{-1/4}
    \left(
    \frac{m_{3/2}}{10\,{\rm GeV}}
    \right)^{1/2}
    \left(
    \frac{\varphi_{\rm osc}}{10^{14}\,{\rm GeV}}
    \right)^{-1} 
        &
        (\varphi_{\rm osc}
        \gtrsim
        \varphi_{\rm cr,N})
        \\[0.6em]
        {\displaystyle
\left(
\frac{N_l}{3}
\right)^{-1/2}
        \left(
        \frac{|K|}{0.1}
        \right)^{1/2}
    \left(
    \frac{T_{\rm R}}{10^5\,{\rm GeV}}
    \right)
    \left(
    \frac{m_{3/2}}{10\,{\rm GeV}}
    \right)^{-1}
    \left(
    \frac{\varphi_{\rm osc}}{10^{14}\,{\rm GeV}}
    \right)^{2} 
    }
        &
        (\varphi_{\rm osc}
        \lesssim
        \varphi_{\rm cr,N}) 
    \end{cases}
    ,
    \label{L nu e in the New type}
\end{align}
where $\varphi_{\rm cr,N}$ is defined as
\begin{align}
    \varphi_{\rm cr,N}
    \equiv
    1.50\times 10^{14}\,{\rm GeV}\,
    \left(
    \frac{N_l}{3}
    \right)^{1/6}
    \left(
    \frac{g_{*,s}}{10.75}
    \right)^{-1/12}
    \left(
    \frac{|K|}{0.1}
    \right)^{-1/6}
    \left(
    \frac{m_{3/2}}{1\,{\rm GeV}}
    \right)^{1/2}
    \left(
    \frac{T_{\rm R}}{10^5\,{\rm GeV}}
    \right)^{-1/3}.
\end{align}.

The resultant baryon-to-entropy ratio from the evaporation process of Q-balls is given by
\begin{align}
    \eta_b
    \simeq
    L_{e}^{\rm init}
    \frac{\Delta Q_b}{Q_{\rm N}^\mathrm{init}},
\end{align}
where $\Delta Q_b$ is the total evaporated baryon charge and is given by
\begin{align}
   &\frac{\Delta Q_b}{Q_{\rm N}^\mathrm{init}}
    \simeq 
    \nonumber \\
    &\begin{cases}
     \displaystyle
        -\frac{48M_{\rm Pl}
        \left[
    -64A\sqrt{|K|}M_{\rm Pl}H(T_{\rm R})m_{3/2}M_s^2
        +
\pi\sqrt{10g_{*}}T_{\rm eq}^3(T_{\rm ew}^2-T_{\rm R}^2)
        +
        192M_{\rm Pl}H(T_{\rm R})T_{\rm eq}^3
        \log
        \left(
        \frac{
        T_{\rm R}
        }{T_{\rm eq}}
        \right)
        \right]
        }
        {23\pi |K|
        g_{*}
        m_{3/2}M_s^2T_{\rm eq}^3}
        &
        (T_{\rm ew} \lesssim T_{\rm R} \lesssim T_{\rm eq})
        \vspace{2mm} 
        \\
        \displaystyle
        -\frac{48M_{\rm Pl}
        \left[
    \sqrt{\frac{10}{g_*}}\pi T_{\rm eq}T_{\rm R}^3(T_{\rm R}^2-T_{\rm ew}^2)
    +2A\sqrt{|K|}m_{3/2}M_s^2
    \left(
    \frac{32}{\sqrt{g_*}}
    M_{\rm Pl}
    H(T_{\rm R})
    T_{\rm eq}
    +\sqrt{10}\pi T_{\rm R}^2
    (T_{\rm R}-T_{\rm eq})
    \right)
        \right]
        }
        {23\pi |K|
        m_{3/2}M_s^2T_{\rm eq} T_{\rm R}^3}
        &
        (T_{\rm ew} \lesssim T_{\rm eq} \lesssim T_{\rm R})
        \vspace{2mm} 
        \\
        \displaystyle
        -\frac{96AM_{\rm Pl}
        \left(
        \pi\sqrt{10g_*}(T_{\rm ew}-T_{\rm R})T_{\rm R}^2
        -32M_{\rm Pl}H(T_{\rm R})
        T_{\rm ew}
        \right)
        }
        {23\pi g_*\sqrt{|K|}
        T_{\rm ew} T_{\rm R}^3}
        &
        (T_{\rm eq}\lesssim T_{\rm ew} \lesssim T_{\rm R})
        \vspace{2mm} 
        \\
        \displaystyle
        -\frac{48M_{\rm Pl}
        \left[
    \sqrt{\frac{10}{g_*}}\pi T_{\rm eq}T_{\rm R}^3(T_{\rm eq}^2-T_{\rm ew}^2)
    +2A\sqrt{|K|}m_{3/2}M_s^2
    \left(
    \frac{32}{\sqrt{g_*}}
    M_{\rm Pl}
    H(T_{\rm R})
    T_{\rm eq}
    +\sqrt{10}\pi T_{\rm R}^2
    (T_{\rm R}-T_{\rm eq})
    \right)
        \right]
        }
        {23\pi |K|
        m_{3/2}M_s^2T_{\rm eq} T_{\rm R}^3}
        &
        (T_{\rm ew} \lesssim T_{\rm eq} \lesssim T_{\rm R})
    \end{cases}.
    \label{The total evaporation charge in the New Type}
\end{align}
Here, we have assumed $T_{\rm eq}<M_s$, which holds in the parameter range of our interest.

\begin{figure}
    \centering
    \includegraphics[width=0.8\linewidth]{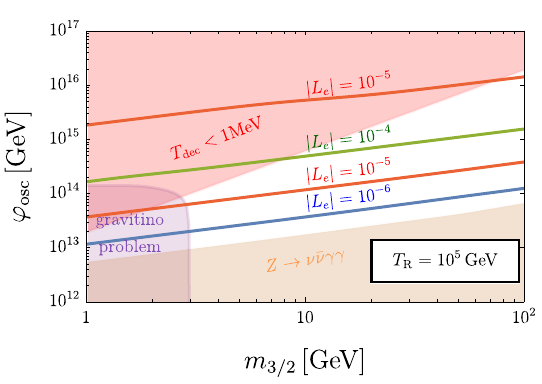}
    \caption{$m_{3/2}$-$\varphi_{\rm osc}$ contour to explain the origin of lepton asymmetry by the decay of New-type Q-balls. }
    \label{fig : new-type case}
\end{figure}

In Fig.~\ref{fig : new-type case}, we show the $m_{3/2}$-$\varphi_{\rm osc}$ contour to explain the origin of lepton asymmetry. Here, we choose the value of $K$ as $K=-0.1$ as a reference value. For this parameter set, in the regime $\varphi_{\rm osc}\gtrsim 10^{14}\,{\rm GeV}$, corresponding to the upper half of the figure, the energy density of Q-balls dominate the total cosmic energy density before the decay. On the other hand, in the regime $\varphi_{\rm osc}\lesssim 10^{14}\,{\rm GeV}$, corresponding to the lower half of the figure, the Q-balls remain subdominant throughout their evolution and decay. The green solid line near the center corresponds to the parameters in which the decay of Q-balls realizes $|L_e|=10^{-4}$. Importantly, both above and below this curve, the final lepton asymmetry is smaller than $|L_e|=10^{-4}$. In the region above the line, the Q-balls begins to dominate the cosmic energy density before their decay. For larger $\varphi_{\rm osc}$, the initial charge $Q_N$ becomes larger, and hence it takes a longer time for them to decay completely. Consequently, the cosmic temperature at the decay of Q-balls, $T_{\rm dec}$, becomes lower, which in turn reduces the resulting lepton-to-entropy ratio (see Eq.~\eqref{eq: L init}). On the other hand, in the region below the line, the Q-balls remain subdominant until their decay. In this regime, smaller $\varphi_{\rm osc}$ leads to a smaller energy fraction of the AD field, $f_{\rm osc}$, and therefore to a smaller lepton-to-entropy ratio. 

The purple shaded region corresponds to the gravitino problem. The scaling of its boundary with respect to $m_{3/2}$ and $\varphi_{\rm osc}$ changes around $\varphi_{\rm osc}\sim 10^{14}\,{\rm GeV}$. This is because, around this value, the cosmological evolution switches between the cases in which the Q-balls do or do not dominate the cosmic energy density before their decay. For $\varphi_{\rm osc}\gtrsim 10^{14}\,{\rm GeV}$, the Q-balls dominate the cosmic energy density before they decay, so that the entropy dilution by the decay of Q-balls affects the final gravitino density parameter. Through this dilution factor, the density parameter depends directly on $\varphi_{\rm osc}$. As $\varphi_{\rm osc}$ increases, the entropy dilution becomes more efficient, and the gravitino problem is alleviated. By contrast, for $\varphi_{\rm osc}\lesssim 10^{14}\,{\rm GeV}$, the Q-balls remain subdominant, and entropy dilution is therefore negligible. In this regime, for fixed $T_{\rm R}$, the gravitino problem is determined solely by $m_{3/2}$, resulting in a boundary parallel to the vertical axis. Finally, let us comment on the orange shaded region. In this region, the decay of Q-balls can generate only $|L_e|\lesssim 10^{-7}$. In that case, explaining the observed isotropic cosmic birefringence angle would require a larger value of $c_\nu/M^2$, which would in turn conflict with the upper bound from rare $Z$-boson decay searches.

Note that, in the New-type case, the baryon asymmetry is typically much suppressed compared to the delayed Q-ball case, alleviating the baryon number overproduction in the parameter region of our interest. 

\subsection{Lepton number fluctuations from Affleck-Dine mechanism}

In this subsection, we discuss the lepton-number fluctuations
generated by inflationary fluctuations of the AD field and the
associated constraints from curvature perturbations.
We consider $c>0$ with $c=\mathcal{O}(1)$. In this regime, the radial
mode has an effective mass of order the Hubble parameter during
inflation, whereas the phase mode is nearly massless because the
A-term is subdominant. We therefore focus on fluctuations in the
phase direction.
The characteristic amplitude of these fluctuations in field space is
\begin{align}
\varphi_{\rm inf}
\sqrt{\langle\delta\theta_{\rm inf}^{2}\rangle}
\sim 
\frac{H_{\rm inf}}{2\sqrt{2}\pi},
\end{align}
where $H_{\rm inf}$ and $\varphi_{\rm inf}$ denote the Hubble parameter
and the radial amplitude of the AD field during inflation,
respectively. 
Since the phase perturbation remains approximately constant until
the A-term becomes important, its amplitude at the onset of coherent
oscillations is estimated as
\begin{align}
\sqrt{\langle\delta\theta_i^{2}\rangle}
&\sim
\frac{1}{2\sqrt{2}\pi}\frac{H_{\rm inf}}{\varphi_{\rm inf}}
\simeq
\frac{1}{2\sqrt{2}\pi}\frac{H_{\rm inf}}{\varphi_{\rm osc}}
\left(
\frac{H_{\rm osc}}{H_{\rm inf}}
\right)^{1/(n-2)}
\nonumber\\
&\simeq
\frac{1}{2\sqrt{2}\pi}\frac{H_{\rm inf}}{\varphi_{\rm osc}}
\left(
\frac{m_{3/2}}{\sqrt{c}H_{\rm inf}}
\right)^{1/(n-2)}.
\label{eq : delta theta i}
\end{align}
Here, $\theta_i$ denotes the phase at the onset of coherent
oscillations. We have used the radial scaling in
Eq.~\eqref{eq : radial profile of AD field right after inf},
together with $H_{\rm osc}\simeq m_{3/2}/\sqrt{c}$.
The inflationary Hubble parameter is constrained by the requirement
that the AD field does not dominate the energy density during inflation.
Using the magnitude of the Hubble-induced mass term in
Eq.~\eqref{AD field potential} to estimate the characteristic
AD-field energy density, we impose the order-of-magnitude condition
\begin{align}
cH_{\rm inf}^{2}\varphi_{\rm inf}^{2}
\lesssim
3M_{\rm Pl}^{2}H_{\rm inf}^{2}.
\label{eq: cond for spectator}
\end{align}
Combining this condition with the radial scaling gives
\begin{align}
H_{\rm inf}
\lesssim
H_{\rm osc}
\left(
\sqrt{\frac{3}{c}}\,
\frac{M_{\rm Pl}}{\varphi_{\rm osc}}
\right)^{n-2}.
\label{eq: cond for spectator 2}
\end{align}
We now consider two consequences of the initial phase fluctuations.
First, they generate lepton-number isocurvature perturbations because
the produced lepton asymmetry depends on the initial phase of the
AD condensate, as shown in
Eq.~\eqref{resulting lepton number in AD leptogenesis}.
For generic initial phases, we estimate the corresponding
power-spectrum amplitude as
\begin{align}
\mathcal{P}_\mu
\sim
\frac{1}{8\pi^2}
\left(
\frac{H_{\rm inf}}{\varphi_{\rm inf}}
\right)^2
\simeq
\frac{1}{8\pi^2}
\left(
\frac{H_{\rm inf}}{\varphi_{\rm osc}}
\right)^2
\left(
\frac{m_{3/2}}{\sqrt{c}H_{\rm inf}}
\right)^{2/(n-2)}.
\label{eq : delta AD}
\end{align}
Combining Eqs.~\eqref{eq: cond for spectator 2}
and~\eqref{eq : delta AD}, and evaluating the resulting bound
for $n=6$ and $c=1$, gives
\begin{align}
\mathcal{P}_\mu
\lesssim
\frac{27M_{\rm Pl}^{6}m_\phi^{2}}
{8\pi^2\varphi_{\rm osc}^{8}},
\label{eq: upper bound of P mu 3}
\end{align}
where we have used $H_{\rm osc}\simeq m_\phi\simeq m_{3/2}$.
We adopt these choices of $n$ and $c$ when evaluating the bounds below. On the other hand, generally, $H_{\rm inf}$ is also bounded by the tensor-to-scalar ratio $r\lesssim 0.03$~\cite{Balkenhol:2025wms,BICEP:2021xfz, Tristram:2021tvh} and its upper bound is roughly given by $H_{\rm inf}\lesssim H_{\rm inf,max}\equiv 4.7\times 10^{13}\,$GeV. Thus, we obtain
\begin{align}
\mathcal{P}_\mu
\lesssim
\min\left[\frac{27M_{\rm Pl}^{6}m_\phi^{2}}
{8\pi^2\varphi_{\rm osc}^{8}},\;
\frac{H_{\rm inf,max}^{3/2}m_{3/2}^{1/2}}{8\pi^2\varphi_{\rm osc}^2}
\right].
\label{eq: upper bound of P mu}
\end{align}
Second, the initial phase fluctuations can also generate energy
density perturbations of the AD condensate. Although their
contribution to the energy density is negligible while the A-term
is subdominant, they can source density perturbations through the
A-term when it becomes important around the onset of coherent
oscillations~\cite{McDonald:2003jk}. The resulting fractional
energy density perturbation is typically estimated as
\begin{align}
\frac{
\sqrt{\langle\delta\rho_{\rm AD}^{2}\rangle}
}{
\bar\rho_{\rm AD}
}
\sim
\frac{1}{2\sqrt{2}\pi}\frac{H_{\rm inf}}{\varphi_{\rm inf}}.
\end{align}
The evolution of the field perturbations and their conversion into
energy density perturbations are discussed in
Appendix~\ref{append: AD perturbation}.
These density perturbations subsequently source curvature
perturbations when the Q-balls decay. Accounting for the fraction
of the total energy density carried by the Q-balls immediately
before their decay, we obtain
\begin{align}
\mathcal{P}_{\zeta,{\rm Q}}
&\sim
\mathcal{P}_\mu
\left(
\frac{f_{\rm osc}}
{f_{\rm osc}+a_{\rm R}/a_{\rm dec}}
\right)^2
\nonumber\\
&\simeq
\frac{1}{8\pi^2}
\left(
\frac{H_{\rm inf}}{\varphi_{\rm osc}}
\right)^2
\left(
\frac{m_{3/2}}{\sqrt{c}H_{\rm inf}}
\right)^{2/(n-2)}
\left(
\frac{f_{\rm osc}}
{f_{\rm osc}+a_{\rm R}/a_{\rm dec}}
\right)^2.
\label{eq : zeta AD}
\end{align}
The last factor is the square of the Q-ball energy fraction
immediately before decay.
To preserve the inflationary predictions for the amplitude and
spectral tilt of the curvature perturbations on CMB scales, we
require
\begin{align}
\mathcal{P}_{\zeta,{\rm Q}}
\ll
\mathcal{P}_\zeta.
\end{align}
Combining the corresponding constraint with
Eq.~\eqref{eq: upper bound of P mu}, we define the following
upper bound for the lepton-number fluctuation:
\begin{align}
\mathcal{P}_\mu
\lesssim
\mathcal{P}_{\mu,{\rm max}}
\equiv
\min\left[
\frac{27M_{\rm Pl}^{6}m_\phi^{2}}
{8\pi^2\varphi_{\rm osc}^{8}}, \;
\frac{H_{\rm inf,max}^{3/2}m_{3/2}^{1/2}}{8\pi^2\varphi_{\rm osc}^2},
\;
\left(
\frac{f_{\rm osc}+a_{\rm R}/a_{\rm dec}}
{f_{\rm osc}}
\right)^2
\mathcal{P}_\zeta
\right].
\label{eq: upper bound of P mu 2}
\end{align}
The first of $\min$ follow from the requirement that
the AD field not dominate during inflation. The second corresponds to the maximum AD field fluctuations allowed by the observational upper bound of $H_{\rm inf}$. The third corresponds
to the boundary at which Q-ball decay would generate curvature
perturbations comparable to the observed ones. Our requirement
$\mathcal{P}_{\zeta,{\rm Q}}\ll\mathcal{P}_\zeta$ therefore
means that $\mathcal{P}_\mu$ must lie well below this third argument.

These estimates apply to both delayed-type and new-type Q-balls.
The ratio $a_{\rm R}/a_{\rm dec}$ is determined using
Eqs.~\eqref{eq : aR over adec} and~\eqref{Qball decay rate},
together with the properties of the relevant Q-ball type.
We discuss the implications for each case below.

\subsubsection{Delayed-type case}

In the delayed-type case, we consider a large initial amplitude,
$\varphi_{\rm osc}\simeq0.1M_{\rm Pl}$. The bound in
Eq.~\eqref{eq: upper bound of P mu} then becomes
\begin{align}
\mathcal{P}_\mu
\lesssim
6\times10^{-30}
\left(
\frac{m_\phi}{1\,{\rm GeV}}
\right)^2
\left(
\frac{\varphi_{\rm osc}}{0.1M_{\rm Pl}}
\right)^{-8}.
\end{align}
For GeV-scale masses, this is far below the observed curvature
power $\mathcal{P}_\zeta$. We therefore neglect the lepton-number
isocurvature perturbations in this case.

\subsubsection{New-type case}

In the new-type case, a smaller value of $\varphi_{\rm osc}$
can substantially weaken the bound in
Eq.~\eqref{eq: upper bound of P mu 2}.
The first and second argument of $\min$ increase rapidly as
$\varphi_{\rm osc}$ decreases, scaling as
$\varphi_{\rm osc}^{-8}$ and $\varphi_{\rm osc}^{-2}$, respectively. The third argument can also increase
if the smaller initial AD-field energy fraction leads to a smaller
Q-ball energy fraction at decay, thereby suppressing the conversion
of AD-field density perturbations into curvature perturbations.
To evaluate this third constraint, we determine
$a_{\rm R}/a_{\rm dec}$ from
Eqs.~\eqref{eq : aR over adec} and~\eqref{Qball decay rate},
using the new-type Q-ball properties given in
Eqs.~\eqref{Q_N}--\eqref{eq :  New type Q-ball prop 4}.
Consequently, the new-type case can accommodate substantially
larger lepton-number fluctuations than the delayed-type case.

\begin{figure}
    \centering
    \includegraphics[width=0.8\linewidth]{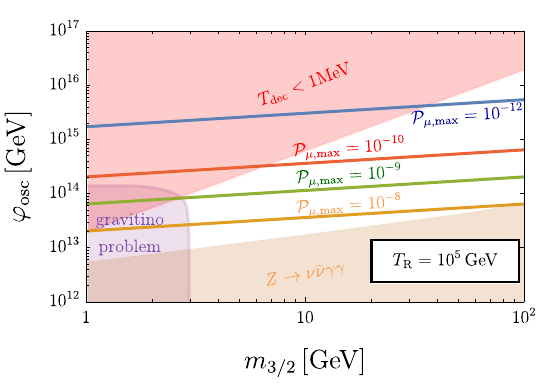}
    \caption{The contours of $\mathcal{P}_{\mu,{\rm max}}$ for new-type Q-balls. The setup and the shaded regions are exactly the same as Fig.~\ref{fig : new-type case}.}
    \label{fig: Lepton isocurvature contour with New type}
\end{figure}
In Fig.~\ref{fig: Lepton isocurvature contour with New type}, we show contours of $\mathcal{P}_{\mu,{\rm max}}$ for new-type Q-balls. Values in the range $10^{-12}\lesssim \mathcal{P}_{\mu,{\rm max}}\lesssim 10^{-7}$ can be realized within the parameter region of interest. As discussed in the preceding paragraph, smaller values of $\varphi_{\rm osc}$ allow larger values of $\mathcal{P}_{\mu,{\rm max}}$. In the parameter region considered in Fig.~\ref{fig: Lepton isocurvature contour with New type}, the second argument in the min in Eq.~\eqref{eq: upper bound of P mu 2} determines the value of $\mathcal{P}_{\mu,{\rm max}}$, which means
\begin{align}
    \mathcal{P}_{\mu,{\rm max}}
    \simeq
    1.3\times 10^{-9}
    \left(
    \frac{H_{\rm inf,max}}{4.7\times 10^{13}\,{\rm GeV}}
    \right)^{3/2}
    \left(
    \frac{m_{3/2}}{10\,{\rm GeV}}
    \right)^{1/2}
    \left(
    \frac{\varphi_{\rm osc}}{10^{14}\,{\rm GeV}}
    \right)^{-2}.
\end{align}

For $\varphi_{\rm osc}\lesssim 10^{14}\,{\rm GeV}$, $\mathcal{P}_{\mu,{\rm max}}$ can reach values larger than $\mathcal{P}_\zeta\simeq 2.1\times 10^{-9}$. In this parameter region where Q-balls remain subdominant in the total energy density until their decay, the resulting curvature power is suppressed by the factor of ${f_{\rm osc}}^2/({f_{\rm osc}+a_{\rm R}/a_{\rm dec}})^2$, as shown in Eq.~\eqref{eq : zeta AD}.
Thus, this region allows lepton-number fluctuations with $\mathcal{P}_{\mu}\gtrsim\mathcal{P}_\zeta$ while maintaining $\mathcal{P}_{\zeta,Q}\ll\mathcal{P}_\zeta$ on CMB scales.

\section{Summary and Discussions}
\label{sec: summary and discussions}
In this paper, we have investigated a vector Chern-Simons coupling between the neutrino current and photons as a possible origin of isotropic cosmic birefringence. In particular, we showed that a large lepton asymmetry generated through the Affleck-Dine mechanism can successfully account for the observed isotropic birefringence angle. Furthermore, we performed the first analysis of anisotropic cosmic birefringence induced by the neutrino current. 

In section~\ref{sec: Model setup}, we reviewed a model of generating cosmic birefringence from cosmic neutrino background via the vector Chern-Simons interactions to photon. In section~\ref{sec: ansotropic CB}, we evaluated the anisotropic cosmic birefringence sourced by lepton isocurvature perturbation. In particular, we focused on the anisotropy generated by the initially scale-invariant isocurvature perturbations of the lepton number. 
We compared the results with numerical integration and the analytical approximation, which is found to be a very good precision with $\ell\lesssim 30$. On the other hand, on small scales with $\ell\gtrsim 30$, the free streaming of the neutrinos and antineutrinos is found to smooth out the lepton number fluctuations, and therefore the perturbations are suppressed more strongly than the analytical prediction.
We also examined the contribution of curvature perturbations and found that it dominates on small scales with $\ell \gtrsim 10^2$.
Intriguingly, compared with the spectrum induced by ALP fluctuations, we find that the suppression is much milder, making the neutrino scenario more promising to probe with CMB observables at high $\ell$.
Our next step is to connect the anisotropic birefringence angular spectrum to CMB observables and evaluate the future detectability of this scenario. 

In section~\ref{sec: AD mechanism as an origin}, as an example we considered Q-ball scenario. Especially, we examined the delayed-type Q-ball scenario and the new-type Q-ball scenario. In the delayed-type case, the resulting lepton asymmetry parameter is typically large, 
$|L_e| \gtrsim 10^{-4}$, allowing the isotropic cosmic birefringence to be realized with a value of 
$c_\nu/M^2$ well below the experimental upper bound. In certain regions of parameter space, 
this case can also account for the recently reported low helium-4 abundance in 
extremely metal-poor galaxies reported by EMPRESS~\cite{Matsumoto:2022tlr,Yanagisawa:2025mgx}. 
However, in this case, Q-balls typically dominate the cosmic energy density before they decay. 
Their fluctuations can therefore significantly contribute to the cosmological curvature 
perturbations. Consequently, in order not to spoil the curvature perturbations on CMB scales, 
a sizable lepton-number isocurvature perturbation, which could otherwise source anisotropic cosmic birefringence, 
cannot be realized.

On the other hand, in the new-type case, the resulting lepton asymmetry parameter can lie in the range
$10^{-7}\lesssim |L_e|\lesssim 10^{-4}$ in the typical parameter region of interest. 
Although this is not sufficient to realize the $1\sigma$-favored value of the lepton asymmetry 
suggested by EMPRESS, it can still account for the isotropic cosmic birefringence without conflicting with the experimental bound from rare $Z$-boson decays. 
In this case, the energy fraction of the AD field is typically much smaller than in the delayed-type case. 
As a result, the AD field fluctuations can generate a sizable lepton-number isocurvature perturbations, which provide a potential source of anisotropic cosmic birefringence, without significantly affecting the curvature perturbations on CMB scales.

In this paper, we have approximated neutrinos as massless throughout the evolution of the universe. At later times, however, when the cosmic temperature becomes comparable to the neutrino mass, gravitational clustering of neutrinos may modify the scale dependence of the anisotropic birefringence angular spectrum. Such an effect could potentially be probed through tomographic approaches to cosmic birefringence, for example by exploiting the signal generated during the cosmic reionization epoch. A more detailed study of this issue is left for future work.

\section*{Acknowledgement}

We would like to thank Patricia Diego-Parazuelos, Yusuke Mikura, Kai Murai and Alexandros Papageorgiou for fruitfull discussion at the online workshop Parity Violation from Home 2025.
We also thank PASCOS 2025 conference at Durham University, where the idea for this work emerged through discussions on the balcony while enjoying the beautiful view of Durham Cathedral.
This work is supported by JSPS KAKENHI Grant Numbers 25KJ1164 (K.K), JST FOREST Program under Contract No. JPMJFR222Y (I.O), JSPS KAKENHI Grant Nos. 19K14702 (I.O) and by KEK support program for young resercher No. B0E510521 (I.O).

\appendix

\section{Effect of the Spatial Component of the Neutrino Current on Anisotropic Cosmic Birefringence}
\label{Append: effect of spatial current on anisotropic CB}

The effect of $J_i$ on birefringence angle (up to the linear order with respect to $J_i$) is given by~\cite{Carroll:1989vb}
\begin{align}
    \delta\beta_{\rm sp}
   =
    \frac{c_\nu}{2M^2}\int {\rm d}\tau\, a(\tau)\hat{\bm{n}}\cdot \bm{J}(\tau, \Delta\tau \hat{\bm{n}}).
    \label{eq: effect of ji on delta beta}
\end{align}
In the following, we show the concrete expressions of Eq.~\eqref{eq: effect of ji on delta beta} using the neutrino dipole moment $\Psi_{\nu,1},\Psi_{\bar{\nu},1}$.
The spatial component of the neutrino current $J_i$ is given by
\begin{align}
    J_i(x)
    =
    \int \frac{{\rm d}^3\bm{q}}{(2\pi)^3a^3}\,
    \frac{\bm{q}}{\epsilon(q)}
    \left(
    f_{\nu}(x,\bm{q})-f_{\bar{\nu}}(x,\bm{q})
    \right).
\end{align}
Thus, by defining the Fourier transform $\tilde{J}(t, \bm{k})$ as
\begin{align}
    \tilde{J}_i(t, \bm{k})
    \equiv
    \int \frac{{\rm d}^3\bm{k}}{(2\pi)^3}e^{i\bm{k}\cdot \bm{x}}\, J_i(t,\bm{x}),
\end{align}
we obtain
\begin{align}
    \tilde{J}_i(t, \bm{k})
    =
    \int \frac{{\rm d}^3\bm{q}}{(2\pi)^3a^3}\,
    \frac{\bm{q}_i}{\epsilon(q)}
    \left(
   f_\nu(\bm{k}, \bm{q},t)-f_{\bar{\nu}}(\bm{k}, \bm{q},t)
    \right).
    \label{eq: ji tilde}
\end{align}
Now by using the expansion
\begin{equation}
f_\nu(\bm{k}, \bm{q},t) = \bar{f}_\nu(q)(1 + \Psi_\nu(\bm{k}, \bm{q},t)) \ ,
\end{equation}
with
\begin{equation}
\Psi_\nu(\bm{k},\bm{q}, \tau) = \sum_{\ell = 0}^\infty (-i)^\ell(2\ell + 1)\Psi_{\nu,\ell}(\bm{k}, q, \tau) P_\ell(\cos\theta) \ ,
\label{eq: Psi nu expansion}
\end{equation}
we obtain
\begin{align}
    \tilde{J}_i(t,\bm{k})
    =
    \sum_{\ell = 0}^\infty (-i)^\ell(2\ell + 1)
    \int{\rm d}\Omega_q
    P_\ell(\hat{\bm{q}}\cdot\hat{\bm{k}})\,
    \hat{\bm{q}}_i
    \int \frac{q^3{\rm d}q}{(2\pi)^3\,\epsilon(q) a^3}
    \left(
    f_{\nu}(q)\Psi_{\nu,\ell}(k, q, \tau)
    -
    f_{\bar\nu}(q)\Psi_{\bar\nu,\ell}(k, q, \tau)
    \right)
    .
\end{align}
Noting that $\Omega_q$ integral survives only when $\ell=1$, we obtain
\begin{align}
    \tilde{J}_i(\tau,\bm{k})
    =
    -i\hat{\bm{k}}_i\mathcal{J}(k,\tau),
    \label{eq: sol of tilde ji}
\end{align}
where we have defined
\begin{align}
    \mathcal{J}(t,\bm{k})
    \equiv
    \frac{g_s}{2\pi^2 a^3}\int \frac{q^3{\rm d}q}{\epsilon(q)}\,
\left(
\bar{f}_{\nu}(q)\Psi_{\nu,1}(k,q,\tau)
    -
    \bar{f}_{\bar{\nu}}(q)\Psi_{{\bar\nu},1}(k,q,\tau)
    \right).
    \label{eq: mathcal J}
\end{align}
Here, we have used the formula
\begin{align}
    \int {\rm d}\Omega_q
    \,\bm{\hat{q}}_i
    \bm{\hat{q}}_j
    =\frac{4\pi}{3}\delta_{ij}.
\end{align}
By combining Eq.~\eqref{eq: effect of ji on delta beta} and the Fourier transform of Eq.~\eqref{eq: sol of tilde ji}, we obtain
\begin{align}
    \delta \beta
    \supset 
    -i\frac{c_\nu}{2M^2}\int {\rm d}\tau\, a(\tau)
    \int \frac{{\rm d}^3\bm{k}}{(2\pi)^3}
    \mu \mathcal{J}(k,\tau)e^{ik\mu\Delta \tau},
\end{align}
with $\mu\equiv \hat{\bm{k}}\cdot\hat{\bm{n}}$. Now multiplying by $Y_{\ell m}(\hat{\bm{n}})$ and performing the angular integral with ${\rm d}\Omega_n$, we obtain
\begin{align}
  \delta\beta(\hat{\bm{n}})
  =\sum b_{\ell m}^{\rm sp}Y_{\ell m}(\hat{\bm{n}}),
\end{align}
where we have defined $b_{\ell m}^{\rm sp}$ as
\begin{align}
    b_{\ell m}^{\rm sp}
    \equiv
    -4\pi i^{\ell}
    \frac{c_\nu}{2M^2}
    \int \frac{{\rm d}^3\bm{k}}{(2\pi)^3}
    Y_{\ell m}^*(\hat{\bm{k}})
    \int {\rm d}\tau a(\tau)\mathcal{J}(k,\tau)j_\ell '(k\Delta\tau).
    \label{eq: b lm sp}
\end{align}
Here, we have used the formula~\footnote{This formula is obtained as follows. By combining Rayleigh's plane-wave formula as
\begin{align}
    e^{ik\mu \chi}
    =
    4\pi\sum_{\ell=0}^\infty
    \sum_{m=-\ell}^\ell
    i^\ell Y_{\ell m}(\hat{\bm{n}})Y_{\ell m}^*(\hat{\bm{k}})j_\ell (k\chi)
\end{align}
and the orthogonality of the spherical harmonics as
\begin{align}
    \int{\rm d}\Omega_n  Y_{\ell m}(\hat{\bm{n}})Y_{\ell' m'}^*(\hat{\bm{n}})=\delta_{\ell \ell'}\delta_{m m'},
\end{align}
we arrive at
\begin{align}
    \int{\rm d}\Omega_n  Y_{\ell m}^*(\hat{\bm{n}})
    e^{ik\mu\chi}=4\pi i^\ell j_\ell(k\chi)
    Y_{\ell m}^*(\hat{\bm{k}}).
    \label{eq: identity}
\end{align}
By differentiating Eq.~\eqref{eq: identity} with $\chi$, we arrive at Eq.~\eqref{eq: identity 2}.
}
\begin{align}
    \int {\rm d}\Omega_n
    Y_{\ell m}^*(\hat{\bm{n}})\mu e^{ik\mu\Delta\tau}
    =4\pi i^{\ell-1}j_{\ell}'(k\Delta\tau)Y_{\ell m}^*(\hat{\bm{k}}).
    \label{eq: identity 2}
\end{align}
Now we evaluate Eq.~\eqref{eq: mathcal J} by primordial values of lepton isocurvature perturbations and curvature perturbations. By approximating that neutrinos are relativistic, the time evolutions of $\Psi_{\nu,1}$ and $\Psi_{\bar\nu,1}$ are given by~\cite{Shoji:2010hm}
\begin{align}
    \tilde{\Psi}_{\nu,1}(k,q,\tau)
    &\simeq
    \tilde{\Psi}_{\nu,0}(k,q,x_i)j_1(x-x_i) +\tilde\Psi_{\nu,1}(k,q,x_i)
    \left(
    j_0(x-x_i)-2j_2(x-x_i)
    \right)
    \nonumber \\
    &+ \tilde\Psi_{\nu,2}(k,q,x_i)
    \left(
    3j_3(x-x_i)-2j_1(x-x_i) 
    \right)
    -\dfrac{{\rm d} \ln \bar{f}_\nu}{{\rm d} \ln q}\psi(k)j_1(x-x_i) 
    \nonumber \\
    &\simeq
    \tilde{\Psi}_{\nu,0}(k,q,x_i)j_1(x-x_i)
    -\dfrac{{\rm d} \ln \bar{f}_\nu}{{\rm d} \ln q}\psi(k)j_1(x-x_i).
    \label{eq: time dependence of dipole}
\end{align}
In the last equality, we have neglected the multipole moment with $\ell\geq 1$ at the initial time. By combining Eqs.~\eqref{eq: mathcal J},~\eqref{eq: b lm sp} and~\eqref{eq: time dependence of dipole}, and in a similar way as Sec.~\ref{sec: ansotropic CB},
we finally obtain
\begin{align}
C^{\beta\beta, {\rm sp}}_\ell &= 
\sum_{x=\mu,\psi}4\pi\left(\dfrac{\beta}{2M^2}\right)^2\int\dfrac{{\rm d}k}{k}\mathcal{P}_{x}(k)\left[\int_e^0
{\rm d}t\,
\bar{n}_L(t)T_{x,{\rm sp}}(k\tau)j_\ell'(k\Delta\tau)\right]^2 \ ,
\label{eq: Cl beta beta for ji}
\end{align}
where we have defined the transfer functions $T_{x,{\rm sp}}(k\tau)$ ($x=\mu,\tau$) as
\begin{align}
    T_{\mu,{\rm sp}}(k\tau)
    &\equiv
    j_1(k\tau),
    \nonumber \\
    T_{\psi,{\rm sp}}(k\tau)
    &\equiv
    2j_1(k\tau).
\end{align}

\begin{figure}
    \centering
    \includegraphics[width=0.75\linewidth]{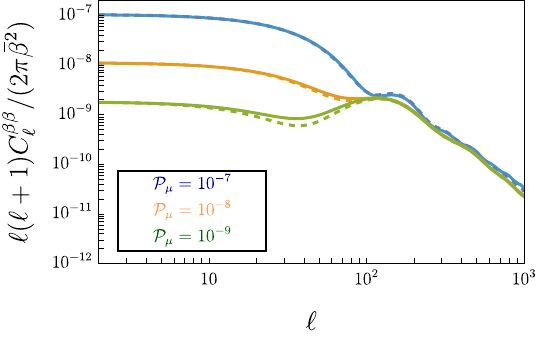}
    \caption{
    Results for the power spectrum of anisotropic cosmic birefringence including the effect of $J_i$, with $\mathcal{P}_\mu=10^{-7} $ (blue), $10^{-8}$ (orange) and $10^{-9}$ (green). The solid lines correspond to $C_\ell ^{\beta\beta,{\rm tot}}$ (Eq.~\eqref{eq: C ell beta beta tot}), while the dashed lines are the results obtained in Sec.~\ref{sec: ansotropic CB}. }
    \label{fig: anisotropic CB including spatial current}
\end{figure}

Also, by combining the $J_i$ effects and $\delta (\Delta n_\nu)$ effects, the total power spectrum is given by
\begin{align}
    C^{\beta\beta, {\rm tot}}_\ell &= 
\sum_{x=\mu,\psi}4\pi\left(\dfrac{\beta}{2M^2}\right)^2\int\dfrac{{\rm d}k}{k}\mathcal{P}_x(k)\left[\int_e^0
{\rm d}t\,
\bar{n}_L(t)
\left(
T_{x}(k\Delta\tau)j_\ell(k\Delta\tau)
-
T_{x,{\rm sp}}(k\Delta\tau)j_\ell'(k\Delta\tau)
\right)
\right]^2 \ .
\label{eq: C ell beta beta tot}
\end{align}
In Fig.~\ref{fig: anisotropic CB including spatial current}, we show the results for the power spectrum of anisotropic cosmic birefringence including the effect of $J_i$, with $\mathcal{P}_\mu=10^{-7},10^{-8},10^{-9}$. The solid lines correspond to $C_\ell ^{\beta\beta,{\rm tot}}$ (Eq.~\eqref{eq: C ell beta beta tot}), while the dashed lines are the results obtained in Sec.~\ref{sec: ansotropic CB}. The two results are in broad agreement, indicating that the effect of the spatial components of the neutrino current on the anisotropic cosmic birefringence is subdominant.

\section{Limber approximation}
\label{append: Limber approximation}
In this Appendix, we explain the approximation used in Eq.~\eqref{eq: Limber approx}. To begin with, we set $\Delta\tau_1\geq\Delta \tau_2$ without loss of generality. Eq.~\eqref{eq: Limber approx} can be recast into an integral form:
\begin{align}
    I_\ell\equiv \int_0^\infty \frac{{\rm d}k}{k}
   j_\ell (k\Delta \tau_1)
   j_\ell (k\Delta \tau_2)
   =
   \frac{1}{2}
   \left(
   \frac{\Delta\tau_2}{\Delta\tau_1}
   \right)^\ell
   \int_0^1 {\rm d}s\, s^{\ell-1}
   \sqrt{(1-s)
   \left(
   1-
   \left(
   \frac{\Delta\tau_2}{\Delta\tau_1}
   \right)^2s
   \right)} \ .
\end{align}
Now by introducing $v\equiv (\ell+1/2)\ln (\Delta\tau_1/\Delta\tau_2)$ and $u\equiv -(\ell+1/2) \ln s$, we obtain
\begin{align}
    I_\ell
    =
    \frac{e^{-v}}{\ell+\frac{1}{2}}\int_0^\infty {\rm d}u\,e^{-u} \sqrt{\sinh \left(
    \frac{u}{2\ell+1}
    \right)
    \sinh \left(
    \frac{u+2v}{2\ell+1}
    \right)
    } \ .
    \label{eq: deform of I ell}
\end{align}
Now we consider the case with $\ell\gg 1$. Because the integrand in Eq.~\eqref{eq: deform of I ell} contains the factor $e^{-u}$, the region with $u\lesssim 1$ is dominant. Thus, we can approximate
\begin{align}
    I_\ell
    \simeq 
    \frac{e^{-v}}{\ell+\frac{1}{2}}\int_0^\infty {\rm d}u\,
    \frac{e^{-u}}{2\ell+1}
    \sqrt{u(u+2v)}
    =
    \frac{vK_1(v)}{2
    \left(\ell+\frac{1}{2}
    \right)^2} \ ,
    \label{eq: deform of I ell 2}
\end{align}
where $K_1$ is the modified Bessel function of second kind. Now with $|\tau_1-\tau_2|\ll \Delta\tau_1\simeq \tau_0$, we can approximate $v\simeq (\ell+1/2)|\tau_1-\tau_2|/\tau_0$. Thus, we obtain
\begin{align}
   \int {\rm d}\tau_1 I_\ell\simeq
   \frac{\tau_0}{
   \left(\ell+\frac{1}{2}
   \right)^3}
   \int_0^\infty
   {\rm d}v\,vK_1(v)
   =
   \frac{\pi\tau_0}{
   2\left(\ell+\frac{1}{2}
   \right)^3} \ ,
   \label{eq: integral of I ell}
\end{align}
for $\ell\gg 1$. 
On the other hand, by using $vK_1(v)\simeq \sqrt{\pi v/2}e^{-v}$ with $v\gtrsim 1$, we see that $I_\ell$ is exponentially suppressed for large $v$. Now the time integral in Eq.~\eqref{eq: C ell beta beta 2} is dominated around $\tau_1\sim \tau_2\sim \tau_i$. Noting that sufficient large $v$ means $|\tau_1-\tau_2|\gg \tau_0/(\ell+1/2)$, we can regard that the only region where $\tau_1\simeq \tau_2$ contributes to the total time integral if $\tau_i\gg \tau_0/(\ell+1/2)$. Therefore, with Eq.~\eqref{eq: integral of I ell}, we arrive at
\begin{align}
   I_\ell\simeq
   \frac{\pi\Delta\tau_1}{
   2\left(\ell+\frac{1}{2}
   \right)^3}
   \delta_D(\tau_1-\tau_2) \ .
   \label{eq: I ell approximation}
\end{align}

\section{Evolution of the superhorizon energy density perturbation of the AD field}
\label{append: AD perturbation}
\begin{figure}
    \centering
    \includegraphics[width=0.75\linewidth]{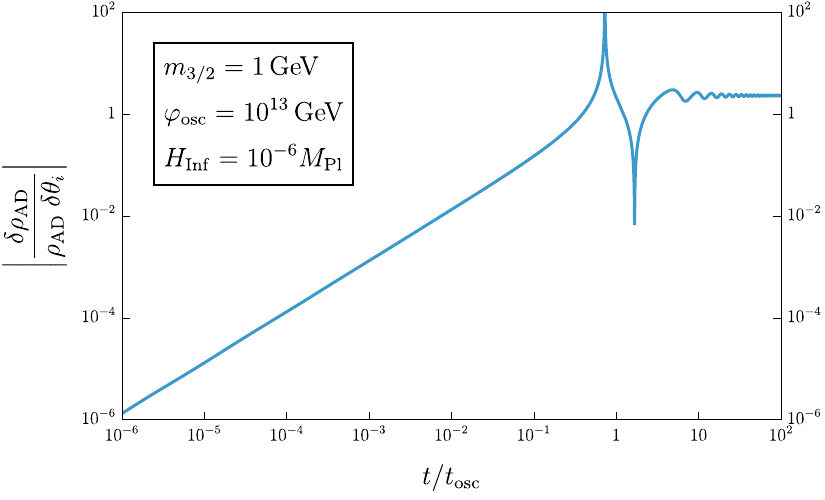}
    \caption{Time evolution of $
\left|
\delta\rho_{\rm AD}
/({\rho_{\rm AD}\,\delta\theta_i})\right| $
for the representative benchmark parameters
$
m_{3/2}=1\,{\rm GeV},
\varphi_{\rm osc}=10^{13}\,{\rm GeV},
H_{\rm Inf}=10^{-6}M_{\rm Pl}
$. The horizontal line is $t/t_{\rm osc}$, with $t_{\rm osc}\simeq m_{3/2}^{-1}$ being the physical time at the onset of the AD field oscillation. We note that, for $t<t_{\rm osc}$, the ratio $\delta\rho_{\rm AD}/\rho_{\rm AD}$ exhibits an apparent divergence at one point. This occurs because the background potential energy of the AD field changes sign as its effective mass transitions from negative to positive.}
    \label{fig: AD field perturbations}
\end{figure}

In this Appendix, we consider the evolution of superhorizon perturbations of the AD field, mainly following Ref.~\cite{McDonald:2003jk}.
We assume that the expansion is dominated by coherent inflaton
oscillations, with
\begin{align}
a(t)\propto t^{2/3},
\qquad
H(t)=\frac{2}{3t}.
\end{align}
We first decompose the AD field as
\begin{align}
\Phi(t,\bm{x})
=
\bigl[\varphi(t)+\delta\varphi(t,\bm{x})\bigr]
e^{i[\theta(t)+\delta\theta(t,\bm{x})]},
\end{align}
where $\varphi$ and $\delta\varphi$ denote the radial background
component and its perturbation, respectively, while $\theta$ and
$\delta\theta$ denote the corresponding phase components.
The background
components satisfy
\begin{align}
\ddot{\varphi}
+\frac{2}{t}\dot{\varphi}
-\varphi\dot{\theta}^{2}
+\frac{1}{2}U_{,\varphi}
+B\varphi^{n-1}\cos\Psi
&=0,
\\
\ddot{\theta}
+\left(
\frac{2}{t}
+2\frac{\dot{\varphi}}{\varphi}
\right)\dot{\theta}
-B\varphi^{n-2}\sin\Psi
&=0.
\end{align}
Here, we have writtn the potential as
\begin{align}
V(\varphi,\theta,t)
&=
U(\varphi,t)
+\frac{2B}{n}\varphi^n\cos\Psi,
\\
U(\varphi,t)
&=
\left(
m_{3/2}^{2}-\frac{4c}{9t^{2}}
\right)\varphi^{2}
+g^{2}\varphi^{2n-2},
\end{align}
and defined
\begin{align}
g
&\equiv
\frac{|\lambda|}{M_{\rm Pl}^{n-3}},
&
B
&\equiv
\frac{|\lambda a_M|m_{3/2}}{M_{\rm Pl}^{n-3}},
&
\Psi
&\equiv
n\theta+\arg(\lambda a_M).
\end{align}
We next consider the evolution of AD-field perturbations on CMB scales.
Since we are interested in the period around the onset of coherent
AD-field oscillations, such scales are superhorizon.
Thus, we can neglect spatial gradient terms.
Linearizing the field equations on the prescribed background gives
\begin{align}
\delta\ddot{\varphi}
+\frac{2}{t}\delta\dot{\varphi}
+
\left[
\frac{1}{2}U_{,\varphi\varphi}
+(n-1)B\varphi^{n-2}\cos\Psi
-\dot{\theta}^{2}
\right]\delta\varphi
-2\varphi\dot{\theta}\,\delta\dot{\theta}
-nB\varphi^{n-1}\sin\Psi\,\delta\theta
&=0,
\nonumber \\
\delta\ddot{\theta}
+\left(
\frac{2}{t}
+2\frac{\dot{\varphi}}{\varphi}
\right)\delta\dot{\theta}
-nB\varphi^{n-2}\cos\Psi\,\delta\theta
+\frac{2\dot{\theta}}{\varphi}\delta\dot{\varphi}
-
\left[
\frac{2\dot{\varphi}\dot{\theta}}{\varphi^{2}}
+(n-2)B\varphi^{n-3}\sin\Psi
\right]\delta\varphi
&=0.
\end{align}
In particular,
\begin{align}
\frac{1}{2}U_{,\varphi\varphi}
=
m_{3/2}^{2}-\frac{4c}{9t^{2}}
+(n-1)(2n-3)g^{2}\varphi^{2n-4}.
\end{align}
We solve the background equations numerically and then use these
solutions to evolve the linear perturbations.
The background energy density of the AD field is
\begin{align}
\rho_{\rm AD}
=
\dot{\varphi}^{2}
+\varphi^{2}\dot{\theta}^{2}
+U(\varphi,t)
+\frac{2B}{n}\varphi^{n}\cos\Psi.
\end{align}
To first order, its perturbation on the same fixed background is
\begin{equation}
\begin{aligned}
\delta\rho_{\rm AD}
={}&
2\dot{\varphi}\,\delta\dot{\varphi}
+2\varphi^{2}\dot{\theta}\,\delta\dot{\theta}
\\
&+
\left[
2\varphi\dot{\theta}^{2}
+U_{,\varphi}
+2B\varphi^{n-1}\cos\Psi
\right]\delta\varphi
\\
&-2B\varphi^{n}\sin\Psi\,\delta\theta.
\end{aligned}
\end{equation}
Fig.~\ref{fig: AD field perturbations} shows the time evolution of $\left|
\delta\rho_{\rm AD}
/({\rho_{\rm AD}\,\delta\theta_i})\right|$
for the representative benchmark parameters
$
m_{3/2}=1\,{\rm GeV},
\varphi_{\rm osc}=10^{13}\,{\rm GeV},
H_{\rm Inf}=10^{-6}M_{\rm Pl}
$.
We assume that the initial perturbations reside in the phase direction,
with $\delta\theta_i$ given by Eq.~\eqref{eq : delta theta i}, and impose
$\delta\varphi(t_i)=\delta\dot{\varphi}(t_i)
=\delta\dot{\theta}(t_i)=0$.
At early times after inflation, the A-term contribution is negligible,
so the initial phase perturbations produce only a negligible
energy density perturbation.
After the AD field starts coherent oscillation, however, the A-term becomes important and allows the initial phase perturbations to source
energy density perturbations.
The value of $\delta\rho_{\rm AD}/\rho_{\rm AD}$ subsequently approaches the same order as $\delta\theta_i$.
For this benchmark, the numerical result therefore supports the
order-of-magnitude estimate
\begin{align}
\left|
\frac{\delta\rho_{\rm AD}}{\rho_{\rm AD}}
\right|
\sim
|\delta\theta_i|,
\end{align}
and hence the estimate used in Eq.~\eqref{eq : zeta AD}.

\bibliographystyle{unsrtnat}
\bibliography{Ref.bib}

\end{document}